\documentclass[9pt,twocolumn,twoside]{osajnl}
\journal{josaa}
\setboolean{shortarticle}{false} 
\usepackage{amsmath,amsfonts,amssymb}
\usepackage{graphicx}
\usepackage[]{aas_macros}

\def\bA{{\bf A}}

\def\bB{{\bf B}}

\def\bc{{\bf c}}

\def\bC{{\bf C}}
\newcommand{\rd}{\mathrm{d}} 
 
\def\bD{{\bf D}}

\def\be{{\bf e}}

\def\bG{{\bf G}}

\def\bH{{\bf H}}

\def\rm{\mathrm{m}}

\def\br{\boldsymbol{r}}

\def\bR{{\bf R}}
\def\bs{\boldsymbol{s}}

\def\rT{\mathrm{T}}

\def\bU{{\bf U}}

\def\bV{{\bf V}}

\def\bW{{\bf W}}

\def\by{{\bf y}}

\def\bnu{\boldsymbol{\nu}}

\def\bSigma{{\bf \Sigma}}

\title{Efficient, Nonlinear Phase Estimation with the Non-Modulated Pyramid Wavefront Sensor}

\author[]{Richard A. Frazin}
\affil[]{ Dept. of Climate and Space Sciences and Engineering, University of Michigan, Ann Arbor, MI 48109} 
\affil[]{E-mail: rfrazin@umich.edu}

\dates{Compiled \today}

\ociscodes{010.1080 Adaptive Optics, 010.7350   Wavefront Sensing}

\doi{\url{http://dx.doi.org/10.1364/ao.XX.XXXXXX}}

\begin{abstract}

The sensitivity of the the pyramid wavefront sensor (PyWFS) has made it a popular choice for astronomical adaptive optics (AAO) systems.
The PyWFS is at its most sensitive when it is used without modulation of the input beam.
In non-modulated mode, the device is highly nonlinear.  Hence, all PyWFS implementations on current AAO systems employ modulation to make the device more linear.
The upcoming era of 30-m class telescopes and the demand for ultra-precise wavefront control stemming from science objectives that include direct imaging of exoplanets make using the PyWFS without modulation desirable. 
This article argues that nonlinear estimation based on Newton's method for nonlinear optimization can be useful for mitigating the effects of nonlinearity in the non-modulated PyWFS.
The proposed approach requires all optical modeling to be pre-computed, which has the advantage of avoiding real-time simulations of beam propagation.   
Further, the required real-time calculations are amenable to massively parallel computation.
Numerical experiments simulate a PyWFS with faces sloped 3.7$^\circ$\ to the horizontal, operating at a wavelength of 0.85 $\mu$m, and with an index of refraction of 1.45.
A singular value analysis shows that the common practice of calculating two "slope" images from the four PyWFS pupil images discards critical information and is unsuitable for the non-modulated PyWFS simulated here.
Instead, this article advocates estimators that use the raw pixel values not only from the four geometrical images of the pupil, but from surrounding pixels as well.
The simulations indicate that nonlinear estimation can be effective when the Strehl ratio of the input beam is greater than 0.3, and the improvement relative to linear estimation tends to increase at larger Strehl ratios.
At Strehl ratios less than about 0.5, the performances of both the nonlinear and linear estimators are relatively insensitive to noise, since they are dominated by nonlinearity error.

\end{abstract}
\setboolean{displaycopyright}{true}
\begin{document} 
\maketitle
\thispagestyle{fancy}

\section{Introduction}\label{sec: intro}

\begin{figure}[t]
\includegraphics[angle=270.,width=0.5\textwidth]{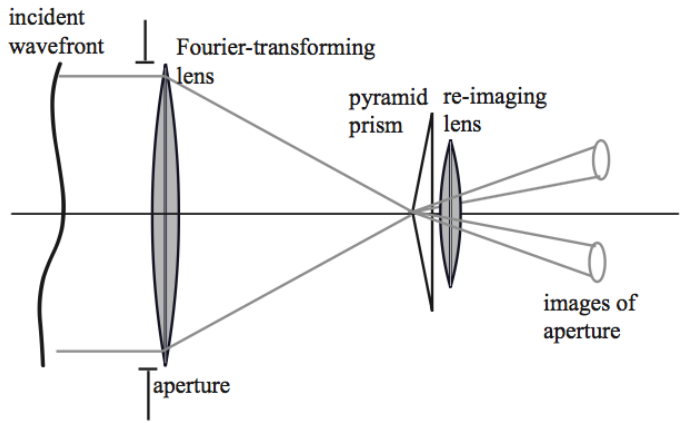}
\caption{\small  Schematic illustration of the PyWFS.  Taken from Ref.~[\citenum{Burvall2006linearity}].}
\label{fig: PyWFS}
\end{figure}

The concept of creating a wavefront sensor for adaptive optics (AO) by focusing a telescope beam onto the vertex of a glass pyramid and then re-imaging the pupil with the light exiting the bottom of the pyramid is due to Ragazzoni.\cite{Ragzzoni_PyWFSinvention_1996}
This has come to be known as the pyramid wavefront sensor (PyWFS), and its design is shown schematically in Fig.~\ref{fig: PyWFS}.
In the standard configuration, the beam passes through a four-sided pyramid, and then a relay lens re-images the pupil, creating four images that carry information about the wavefront.
Usually, the pyramid is a square pyramid, but other configurations have been considered.\cite{Wang_2vs4sidedPy_OE10, Fauvarque_FlatPyramid, Fauvarque_JATIS17, VanKooten_alternativePy_JATIS17}
The desirability of the PyWFS is due to the fact that it is much more sensitive (in terms of performance for a given guide-star magnitude) than the Shack-Hartmann array for low-order modes, and for high-order modes, it is equally sensitive.\cite{Ragazzoni_sensitivity1999, Esposito_Riccardi_AA2001, Chew_PySH_comparison_OptComm06, Viotto_PyramidBehavior2013}

The PyWFS' current status as the most promising and powerful type of wavefront sensor for AO was achieved with release of the unprecedented results produced by the  First Light AO system on the Large Binocular Telescope, which routinely delivers Strehl ratios greater than 0.8 in the H-band.\cite{FLAO_LBT2010, Esposito_FLAOstatus2012}
The PyWFS has become the wavefront sensor of choice in much of the AO community, and is currently employed by MagAO, SCExAO, and the First Light Adaptive Optics (FLAO) system, among others.\cite{MagAO2014, SCExAO_PASP15, Esposito_FLAOstatus2012}
Development of PyWFS technology continues to be high priority research, as the European Extremely Large Telescope (E-ELT), the Giant Magellan Telescope (GMT) and the Thirty Meter Telescope (TMT), all members of the coming generation of 30-meter class telescopes, plan to utilize the PyWFS for their AO programs.\cite{ElHadi_EELT_PyWFS2013, Bernstein_GMToverview2014, BoyerAOatTMT2015}

A necessary consequence of the high sensitivity of the PyWFS is nonlinearity, meaning the intensity seen by the PyWFS detector is nonlinear in the phase of the wavefront.
In order to reduce the nonlinearity, all current implementations of the PyWFS use dynamic modulation of the input beam, in which the focal point is steered around the tip of the pyramid with a period at least several times shorter than the exposure time.\cite{Ragzzoni_PyWFSinvention_1996}
This practice indeed improves the linearity range of the device, but at the expense of sensitivity.\cite{Verinaud2004PyWFS, Fauvarque_FourierFormalism}
In the context of multi-conjugate AO, Ref.~\cite{Greggio_VeryLinearPy_13} proposes the Very Linear Wavefront Sensor, which operates on a locally closed loop to bring the signal into the linear regime.
Ref.~\cite{Viotto_PyLocalLinearity_SPIE16} proposes treating nonlinearity by measuring the nonlinear response to various wavefront modes deliberately imposed on the beam via DM commands.
This scheme cannot treat the fact that the nonlinear response will depend on the instantaneous shape of the wavefront and this effect must be negligible for it to be successful. 
We note that several groups have proposed using spatial light modulators in lieu of mechanical beam steering.\cite{Wang_Py_SLM_OE11, Akondi_Py_SLM_13}

The loss in sensitivity caused by the modulation may not be acceptable for the needs of  so-called "extreme AO" (ExAO).\cite{Guyon_Limits_ApJ2005}
That is why several groups, including the GMT team, hope to use the PyWFS without modulation, using a scheme in which the control loop is first closed while employing modulation, and then the modulation would be turned off.\cite{Males_PathExAO2016}
In these applications, it is likely that the PyWFS will receive a beam that has been partially corrected by atmospheric distortion by a first-stage AO system.
The two-stage AO scheme is currently employed by several extreme AO systems including SCExAO, which takes the beam after it has passed through the Subaru telescope's AO188 system.\cite{SCExAO_PASP15}
This article argues that nonlinear estimation of the phase of the wavefront measured by the non-modulated PyWFS has the potential to be a useful component of such a scheme for ExAO, and that it is worthy of more detailed study.

The PyWFS is intractable analytically, and analytical results have relied on various simplifications and assumptions, including treating the pyramid faces as Foucault knife-edges and neglecting interference between the pupil images [e.g., \citenum{Verinaud2004PyWFS, Burvall2006linearity, Chew_PySH_comparison_OptComm06}].
Perhaps as a result of such studies, the PyWFS is often considered to be essentially a slope-sensor that can be treated as a Shack-Hartmann array, once the sum of one pair of images has been subtracted from the sum of the other pair.\cite{Burvall2006linearity} 
Later literature has discussed the fact that PyWFS is not simply a slope sensor.\cite{Wang_2vs4sidedPy_OE10, Fauvarque_JATIS17}

The first attempt (that we know of) to treat the nonlinearity in the PyWFS computationally was published by Korkiakoski et al.\cite{KorkiPyWFSreconNonlin2007}
That paper is similar to the one here in that a forward model of the PyWFS based on Fourier optics is used to find a relationship between the measurements and the unknown phases, but it neglects interference between the beams exiting the pyramid; this interference effect is not negligible for the simulations presented below.
The nonlinear estimation method in that article utilizes only the gradients of the PyWFS intensities, whereas one of the algorithms in this article utilizes a more powerful nonlinear solution method enabled by calculation of the Hessians of the intensities.
Crucially, the paper by Korkiakoski et al. does not show how the gradient computations can be dramatically accelerated by pre-computation of the modeling calculations and efficient representation of the field in the pupil plane, as is done here.

Ref.~\cite{Shato_PyrRecon2017} provides a comprehensive review of reconstruction methods for the PyWFS and it also introduces two new reconstruction algorithms.
The standard method for reconstruction is called matrix-vector multiplication, in which one pre-computes the (possibly regularized) inverse of a matrix that maps the deformable mirror (DM) commands to the intensity changes.
All of the current reconstruction methods use a linear relationship between the sought-after quantities (i.e., phases or deformable mirror commands) and the intensity measurements.
Almost all (closed-loop) implementations of the PyWFS employ a scheme in which the DM and PyWFS are calibrated only in reference to each other, i.e., the calibration measurements use intensity changes seen by the PyWFS detector in response to known DM commands.
In contrast, the methods described here are enabled by a pre-computed and independently calibrated model of the PyWFS.
One advantage of using such a pre-computed model is that no propagation computations are performed in real time, allowing time-consuming numerical methods for optical modeling to be employed.
Specifically, we will show that the gradients and/or Hessians required by standard methods for nonlinear optimization can be implemented efficiently in a massively parallel fashion, making deployment on real-time AO control systems a realistic possibility.

\section{Fourier Optics Model}\label{sec: model}

One crucial feature of the wavefront estimation method presented here is that it can work with any numerical model used to simulate the pyramid wavefront sensor, with no consequences for the amount of real-time computation required.
Indeed, the validity of the estimation method relies only on the linear relationship between the electric field at the detector and the electric field in the pupil plane, and this linearity will hold for any wavefront sensor so-far ever considered for AO.
(Nonlinear devices, such as optical parametric amplifiers, would violate the linearity assumption.)  
The numerical model employed here for the purpose of algorithm development is based on Fourier optics.\cite{IntroFourierOptics}
In these simulations, the lenses are treated as thin phase-screens that act mathematically as Fourier transform operators.
The required transforms are implemented numerically with discrete Fourier transforms (via an FFT algorithm).
A Fourier optics treatment of the PyWFS was first published in Ref.~\cite{KorkiPyWFSreconNonlin2007} and was later placed into a generalized framework in Refs.~\cite{Fauvarque_FourierFormalism, Fauvarque_JATIS17}.
A sophisticated ray-tracing tool for the PyWFS was presented in Ref.~\cite{Antichi_PyramidRayTrace_JATIS16}.

\begin{figure}[t]
\begin{tabular}{l}
\includegraphics[height=60mm]{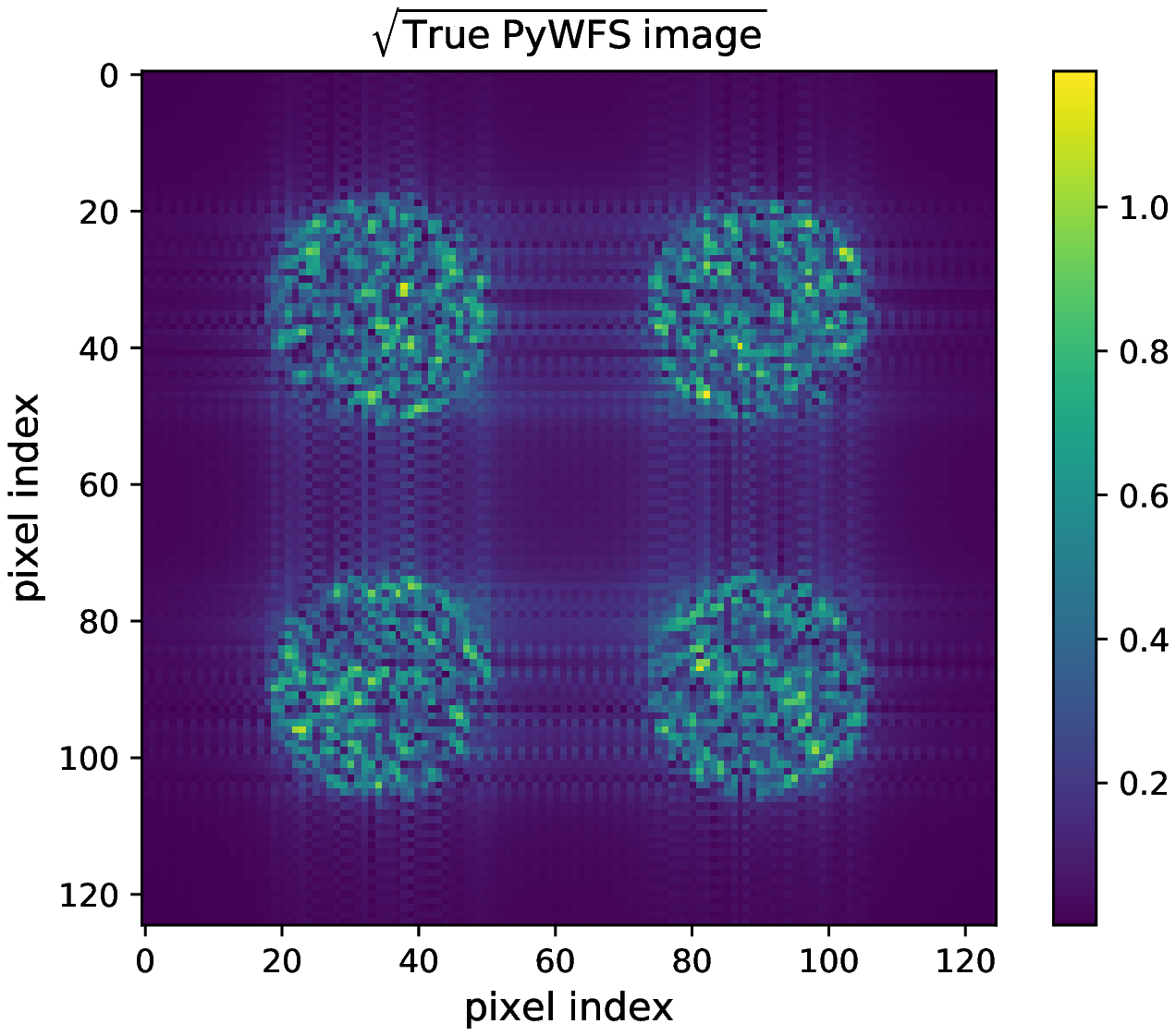} \\
\includegraphics[height=60mm]{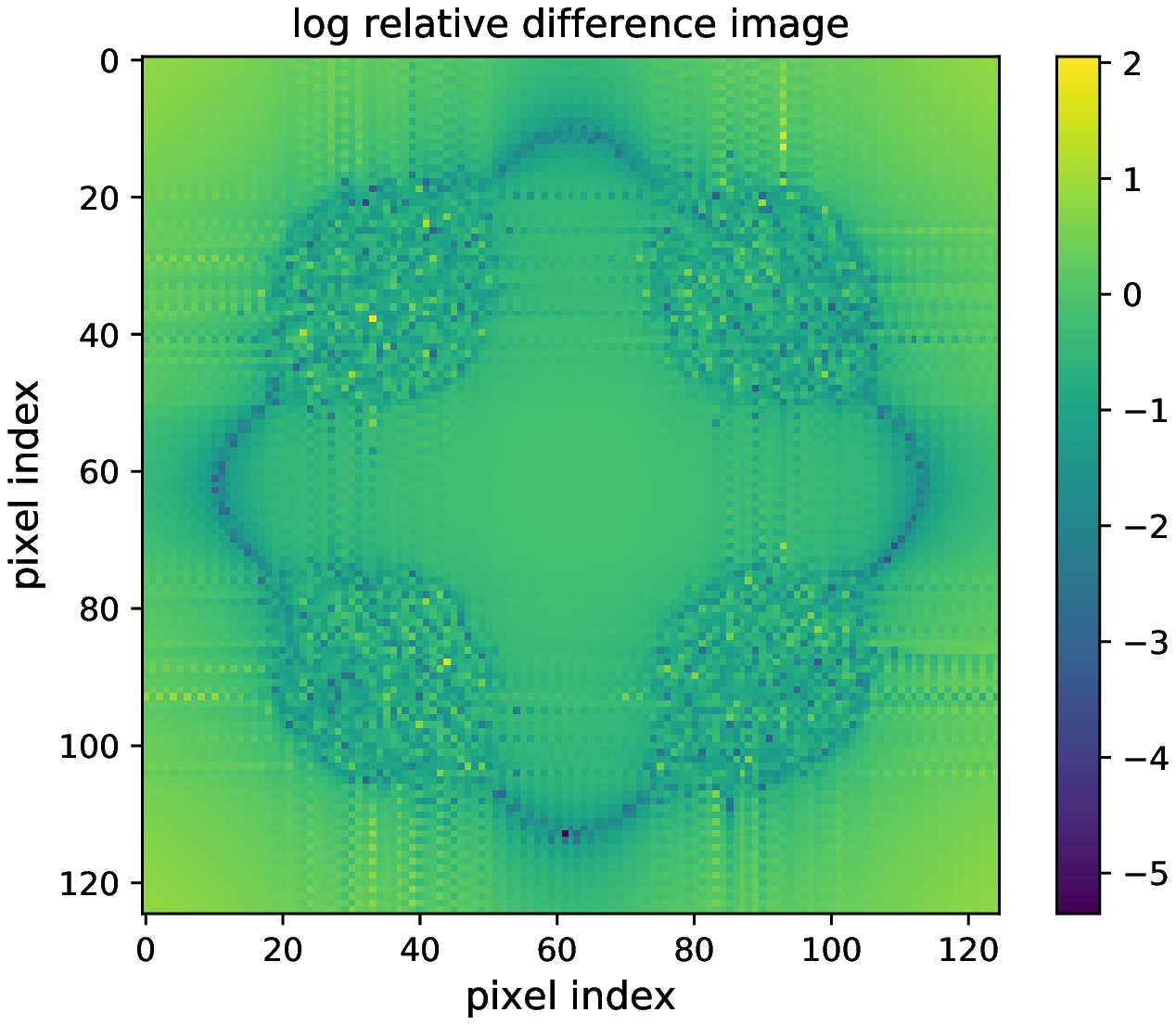}
\end{tabular}
\caption{\small Examples from the PyWFS simulations for a wavefront with a standard deviation 1.1 radians, corresponding to a Strehl ratio of about 0.3.
Each image shows the central $125 \times 125$ pixels of the detector plane.
\emph{top:} $\sqrt{\mathrm{intensity}}$ (normalized units) image.
\emph{bottom:} A representation of the error due to neglecting interference between the beams in the simulation.
The quantity displayed is $\log_{10}[ |I_\mathrm{t} - I_\mathrm{a}| / I_\mathrm{t}]$, where $I_\mathrm{t}$\ is the true intensity calculated by the full simulation, and $I_\mathrm{a}$ is an approximate intensity obtained by neglecting interference between the beams.
Off of the four pupil images, one can see the effects of aliasing in the FFT-based propagation (as explained in the text, more sophisticated modeling is not a barrier to on-sky implementation). 
}
\label{fig: PyramidImages}
\end{figure}

Consider quasi-monochromatic light at wavelength $\lambda$ (wavenumber $k = 2\pi/ \lambda$) impinging on the PyWFS entrance pupil $\mathcal{A}$, and let the electric field in that pupil be represented by $u_0(\br)$ where $\br$ is the two-dimensional (2D) spatial coordinate.
This complex-valued field can be written in terms of its real-valued amplitude $a_0(\br)$ and real-valued phase $\phi_0(\br)$ as $u_0(\br) = a_0(\br) \exp [j \phi_0(\br)]$.
In a PyWFS, the entrance pupil light is brought to a focus with a lens of focal length $f$, and the field in the focal plane $u_f(\br)$, apart from an inconsequential phase factor and a multiplicative constant, is given by \cite{IntroFourierOptics}:
\begin{equation}
u_f(\br) = \int_\mathcal{A} \rd \br' \, u_0(\br') \exp j \left[ -\frac{k}{f} \br' \cdot \br  \right]  \, .
\label{eq: FirstFT}
\end{equation}
The slope of the pyramid faces imparts a spatially variant phase shift to the field in the focal plane.
Let us call this effect the \emph{phase ramp}, denoted as $\mathrm{ramp}_i(\br)$, where $i$ is the face index [so, for a square pyramid $i = (0, \, 1, \, 2, \, 3)$], and it is given by 
\begin{equation}
\mathrm{ramp}_i(\br) = \exp \left[ -j 2\pi(n-1)\frac{\zeta_i(\br)}{\lambda} \right] \, ,
\label{eq: Ramp}
\end{equation}
where $n$\ is the index of refraction of the pyramid glass, and $\zeta_i(\br) \, \leq 0$ is the height of the $i$\underline{th} pyramid face relative to the horizontal.
After the phase ramp is applied, the light propagates through the rest of the pyramid, exits the prism's bottom, and then encounters the collimator lens.
The collimator lens re-images the pupil, but now there is one image of the pupil corresponding to each face of the pyramid (assuming the slope of the pyramid is great enough to separate them; see Ref.~[\citenum{Fauvarque_FlatPyramid}] for a counter-example).
The collimator lens essentially takes a second Fourier transform of the light, and the field on the detector is given by:
\begin{equation}
u_d(\bs) =  \int \rd \br \, \mathrm{ramp}(\br)  u_f(\br)  \exp j \left[ -\frac{k}{f} \br \cdot \bs  \right]  \,
\label{eq: SecondFT}
\end{equation}
 where $\bs$ is the 2D spatial coordinate in the detector plane, $\mathrm{ramp}(\br')$ 
(without the index subscript) is the collection of all of the pyramid faces, and the focal length of the collimator lens is also taken to be $f$.
The spatial limits of integration are taken to be infinite (since the field in the focal plane is essentially contained within any physical boundary of the apparatus).
Combining Eqs.~(\ref{eq: FirstFT}) and (\ref{eq: SecondFT}) gives an expression for the \emph{pyramid operator} $P$:
\begin{align}
P\big(u_0(\br), \bs \big) =  \int & \rd \br' \, \mathrm{ramp}(\br')   \exp j \left[ -\frac{k}{f} \br' \cdot \bs  \right] 
\nonumber \\ & \times
 \int_\mathcal{A} \rd \br \,  u_0(\br) \exp j \left[ -\frac{k}{f} \br \cdot \br'  \right] \, .
\label{eq: PyramidOp}
\end{align}
Note that the pyramid operator is linear in $u_0(\br)$, a fact that will prove critical in designing efficient estimation algorithms.
Eq.~(\ref{eq: PyramidOp}) shows that the pyramid operator essentially takes two Fourier transforms of the pupil field (applying the phase ramp in between), thereby creating inverted images of the pupil.
Then, the field impinging on the detector is conveniently re-expressed as:
\begin{equation}
u_d(\bs) = P\big(u_0(\br), \bs \big)
\label{eq: DetectorField}
\end{equation}
The intensity of the light on the detector surface is:
\begin{equation}
I_d(\bs) = u_d(\bs)u_d^*(\bs) = P\big(u_0(\br), \bs \big)  P^*\big(u_0(\br), \bs \big) \, ,
\label{eq: Intensity}
\end{equation} 
where the superscript $^*$\ signifies complex conjugation.
The goal of wavefront sensing is, of course, to obtain information about the pupil field $u_0(\br)$\ from the measurements of the intensity $I_d(\bs)$.
The sought-after information is usually the phase $\phi_0(\br)$, but the device is also sensitive to the amplitude $a_0(\br)$, which also can be estimated under favorable conditions.

\section{Discretization of the Model}\label{sec: discretization}

Any numerical model must relate the pupil field $u_0(\br)$ at a discrete set of points $\{\br_k\}$, $0 \leq k < K - 1$ to the detector field $u_d(\bs)$ at another discrete set of points $\{\bs_l\}$, $0 \leq l < L - 1$, where $\{\br_k\}$\ and $\{\bs_l\}$ are the sets of all such points in the model and $K$ and $L$ are the numbers of points in each set.
Thus, the detector field at the point $\bs_l$, $u_d(\bs_l)$, can be considered to be a function of the pupil field at each point in the set $\{\br_k\}$, and, using the pyramid operator Eq.~(\ref{eq: PyramidOp}),  this can be expressed as:
\begin{equation}
u_d(\bs_l) = P\big(u_0(\br_0), \, \dots, \, u_0(\br_{K-1}) ; \, \bs_l \big) \, ,
\label{eq: DiscretePyramidOp}
\end{equation}
where the pyramid operator has been taken to have a discrete implementation via some numerical model.
It is useful to define a basis set on the points in the pupil $\{\br_k\}$.
The basis vectors $\{ \be_k \}$,  $0 \leq k < K - 1$ are defined as:
\begin{equation}
\be_k  = (0, \, \dots, \, 0, \, 1, \, 0, \, \dots, \, 0  )   
\label{eq: BasisDef}
\end{equation}
where only the $k$\underline{th} of $K$ entries is nonzero.
Then, via the linearity property of the pyramid operator, Eq.~(\ref{eq: DiscretePyramidOp}) can be written in terms of the basis vectors $\{\be_k\}$ as:
\begin{equation}
u_d(\bs_l) = \sum_{k=0}^{K-1} u_0(\br_k) P\big( \be_k ; \, \bs_l \big) \, .
\label{eq: PyramidBasis}
\end{equation}
Eq.~(\ref{eq: PyramidBasis}) shows that once the pyramid operator has been evaluated for each of the basis vectors $\{ \be_k \}$, the detector field can be calculated with very few operations, independently of the complexity of the optical model used to calculate the pyramid operator.
Indeed, $ \{ P\big( \be_k ; \, \bs_l \big) \}$\ can be stored as a $L \times K$ matrix, and then Eq.~(\ref{eq: PyramidBasis}) can be carried out with a single (complex-valued) matrix-vector multiplication for any desired entrance pupil field represented as $\{ u_0(\br_k) \}$.
For the purposes of discussing the algorithms in the remainder of this article, it will be assumed that the values of  $ \{ P\big( \be_k ; \, \bs_l \big) \}$ have been stored and therefore require no computation.

\subsection{Pupil Field Representations}

We have not specified how the pupil field $\{ u_0(\br_k) \}$\ will itself be represented, and below we present several options.
Each representation expresses the set of field values $\{ u_0(\br_k) \}$\ in terms of $N$\ real-valued parameters $\{c_n\}$.
The first option is called the "PolarPixel" representation:
\begin{equation}
\overset{\mathrm{PolarPixel}}{
u_0(\br_k)} = c_{k + K} \exp[ j (c_k)    ] \: , \;   0 \leq k < K - 1 \, ,
\label{PolarPixel}
\end{equation}
in which $c_0$ through $c_{K-1}$\ represent the phases of the pupil field at each point in the set $\{ \br_k \}$\ and $c_K$ through $c_{2K-1}$\ represent the amplitudes, so  the number of parameters is twice the number of pupil plane pixels, i.e., $N = 2K$.
If the amplitude of the field is known (or perhaps measured by a scintillation monitor), or amplitude effects are ignored, then Eq.~(\ref{PolarPixel}) can be simplified to produce the "PhasePixel" representation:
\begin{equation}
\overset{\mathrm{PhasePixel}}{
u_0(\br_k)} = a_k \exp[ j (c_k)    ] \: , \;   0 \leq k < K - 1 \, ,
\label{PhasePixel}
\end{equation}
where the (positive, real-valued) amplitudes $a_k$\ are assumed to be known.
If amplitude effects are ignored, then the $a_k$\ can be set to the expected value of the magnitude of the pupil field, which can be estimated by summing up all of the counts on the detector to get a mean modulus of the pupil field and applying the appropriate calibrations.
In the PhasePixel representation $N=K$ (so, there is one parameter for each pupil plane pixel).
Another representation with similar efficiency advantages (to be described later) is called the "ReImPixel" representation, in which one models the real and imaginary parts of the field directly:
\begin{equation}
\overset{\mathrm{ReImPixel}}{
u_0(\br_k) } =  c_k + jc_{k + K} \: , \;   0 \leq k < K - 1 \, ,
\label{ReImPixel} 
\end{equation}
where the number of parameters is twice the number of pupil plane pixels, i.e., $N = 2K$.
Thus, in the ReImPixel representation, the amplitude of the pupil plane field at point $k$\ is given by $\sqrt{c_k^2 + c_{k+K}^2}$\ and its phase is given by $\tan^{-1}( c_{k+K}/ c_k )$.

In contrast to the pixel basis functions above, consider a modal representation of the phase (ignoring amplitude effects for simplicity), called "ModalPhase." 
In the ModalPhase representation, the phase is taken to be a linear combination of $N$ modal functions $\{ \psi_n(\br) \}$, such as Zernike polynomials, and the pupil plane field is given by:
\begin{equation}
\overset{\mathrm{ModalPhase}}{
 u_0(\br_k)}  =  a_k \exp j \left[ \sum_{n=0}^{N-1} c_n \psi_n(\br_k)  \right] \,
\label{eq: ModalPhase}
\end{equation}
where the amplitudes $\{ a_k \}$ are taken to be known as in the PhasePixel representation.
Below, we will see that the ModalPhase representation carries a significant computational cost burden that the other representations described above do not share.

\subsection{Detector Intensity and its Derivatives}\label{sec: DI and D}

The estimation algorithms will require expressions for the derivatives of the intensity in the detector plane with respect to the parameters $\{ c_k \}$.
The first step in calculating the intensity derivatives is to calculate the "field Jacobian," which is the derivative with respect to the parameters $\{ c_k \}$ of the complex-valued electric field in the detector plane.
Unless otherwise stated, all formulae given in the remainder of this article assume the PhasePixel representation, but corresponding expressions for the other representations can be derived similarly.
The field and the field Jacobian are given by Eqs.~ (\ref{eq: PyramidBasis}) and (\ref{PhasePixel}), resulting in:
\begin{eqnarray}
 u_d(\bs_l; \bc) & =  & \sum_{k=0}^{K-1} a_k e^{jc_k} P\big( \be_k ; \, \bs_l \big) \, , \; \mathrm{and}
\label{eq: PhasePixelDetField} \\
\frac{\partial}{\partial c_m} u_d(\bs_l; \bc) & = & j a_m e^{jc_m} P\big( \be_m ; \, \bs_l \big) \, .
\label{eq: PhasePixelDetFieldD1}
\end{eqnarray}
Thus, the field Jacobian can be calculated with only several times $LN$\ floating-point operations (FLOPs), where the reader will recall that $K=N$\ for the PhasePixel representation and the $\{a_k \}$\ are taken to be known amplitudes.
The most efficient way to calculate the field values in Eq.~(\ref{eq: PhasePixelDetField}) is to calculate the field Jacobian from Eq.~(\ref{eq: PhasePixelDetFieldD1}) first.
Then, for each position $\bs_l$, sum the corresponding $N$\ values of the gradient (multiplied by $-j$). 
Since elements of the field Jacobian can be calculated completely independently from each other, this can be implemented in parallel.
The field Hessian, i.e., the second derivative matrix of the field at a given point in the detector plane $\bs_l$, has a convenient property that follows directly from Eq.~(\ref{eq: PhasePixelDetFieldD1}):
\begin{equation}
\frac{\partial^2}{\partial c_n \partial c_m} u_d(\bs_l; \bc) = 0  \: \: \:  (m \neq n) \, .
\label{eq: PhasePixelDetFieldD2}
\end{equation}

In order to appreciate the computational simplicity of the PhasePixel representation compared to the ModalPhase representation, consider the expressions equivalent to Eqs. (\ref{eq: PhasePixelDetField})~and~(\ref{eq: PhasePixelDetFieldD1}) under the ModalPhase representation:
\begin{align}
\overset{\mathrm{ModalPhase}}{u_d(\bs_l; \bc)}
 =   \sum_{k=0}^{K-1} a_k & P\big( \be_k ; \, \bs_l \big)  \; \exp j \left[ \sum_{n=0}^{N-1} c_n \psi_n(\br_k)  \right] 
\label{eq: ModalPhaseDetField} \\
\overset{\mathrm{ModalPhase}}{\frac{\partial}{\partial c_m} u_d(\bs_l; \bc) }
 =  j \sum_{k=0}^{K-1}   a_k & P\big( \be_k ; \, \bs_l \big)  \nonumber \\ 
& \times \psi_m(\br_k)
  \; \exp j \left[ \sum_{n=0}^{N-1} c_n \psi_n(\br_k)  \right] \, .
\label{eq: ModalPhaseDetFieldD1}
\end{align}
In particular, one should note the double summation in \eqref{eq: ModalPhaseDetFieldD1}, as opposed to no summations in \eqref{eq: PhasePixelDetFieldD1}.
Unlike the PhasePixel representation [see \eqref{eq: PhasePixelDetFieldD2}], the off-diagonal elements of the field Hessian under the ModalPhase representation are non-zero, and their cumbersome formula is not given here.

Next, we derive expressions for the derivatives of the intensity in the detector plane with respect to the unknown parameters $\{ c_k \}$ under the PhasePixel representation.
The intensity at the detector is given by $ u_d(\bs_l; \bc) u_d^*(\bs_l; \bc)$, and the intensity Jacobian is calculated easily with the help of Eq.~(\ref{eq: PhasePixelDetFieldD1}):
\begin{equation}
\frac{\partial}{\partial c_m} I_d(\bs_l; \bc) =  \left[ \frac{\partial}{\partial c_m} u_d(\bs_l; \bc) \right] u_d^*(\bs_l; \bc) 
+ \: \mathrm{c.c.} \, ,
\label{eq: PhasePixelIntensityD1}
\end{equation}
where "c.c." indicates the complex conjugate of all preceding terms.
Note that for any quantity $h$, $h + \mathrm{c.c.} = 2\Re(h)$, where $\Re(h)$\ is the real part of $h$ [the imaginary part of $h$ is denoted by $\Im(h)$].
This shows that calculating the $L \times M$\ intensity Jacobian requires little more computation, once the field Jacobian has been calculated from Eq.~(\ref{eq: PhasePixelDetFieldD1}).
The Hessian matrix of the intensity at the point $\bs_k$ is given by differentiating Eq.~(\ref{eq: PhasePixelIntensityD1}) with the help of Eqs.~(\ref{eq: PhasePixelDetField}) through (\ref{eq: PhasePixelDetFieldD2}):
\begin{align} 
\frac{\partial^2}{\partial c_n \partial c_m} I_d(\bs_l; \bc) = & 
  \left[ \frac{\partial}{\partial c_m} u_d(\bs_l; \bc) \right]   \left[ \frac{\partial}{\partial c_n} u_d(\bs_l; \bc) \right]^*
 \nonumber \\ 
 & \: \: + \mathrm{c.c.} \: \: \: (m \neq n) 
\label{eq: PhasePixelIntensityD2} \\
\frac{\partial^2}{\partial c_m^2}I_d(\bs_l; \bc)  = &
\left| \frac{\partial}{\partial c_m} u_d(\bs_l; \bc) \right|^2 + \left[ \frac{\partial^2}{\partial c_m^2} u_d(\bs_l; \bc) \right] 
 u_d^*(\bs_l; \bc) 
 \: 
 \nonumber \\
& \: \: + \: \mathrm{c.c.}
\label{eq: PhasePixelIntensityD2diag}
\end{align}
Note that the intensity Hessian in Eqs.~(\ref{eq: PhasePixelIntensityD2}) and (\ref{eq: PhasePixelIntensityD2diag}) is calculated with little more effort, given the already determined values of the field Jacobian in Eq.~(\ref{eq: PhasePixelDetFieldD1}).
Each of the $L$ detector plane positions $\{ \bs_k \}$\ has a $N \times N$ Hessian matrix associated with it, each of which has $(N^2 + N)/2$ unique elements (due to symmetry), for a total of $ L(N^2 + N)/2$\ elements needed to specify the Hessian at every position in the detector plane.
According to Eq.~(\ref{eq: PhasePixelIntensityD2}), each of these $ L(N^2 + N)/2$\ quantities can be calculated with only a few FLOPs, and it is clear that this task can be achieved in a massively parallel manner, since the computation of these elements can be performed independently.
It is worth remarking that much more computation would be required to calculate the Hessian in the ModalPhase representation for a comparable value of $N$, as can be seen by Eq.~(\ref{eq: ModalPhaseDetField}) and considering the steps needed to derive the equations corresponding to Eqs.~(\ref{eq: PhasePixelIntensityD1}) through (\ref{eq: PhasePixelIntensityD2diag}).
In contrast, the PolarPixel and ReImPixel representations have $N=2K$ (instead of $N=K$), but otherwise have similar computational complexity to the PhasePixel representation.

\subsection{Effective Sparsity of the Hessian Matrices}\label{sec: sparsity}

Nonlinear estimation based on Newton's method for optimization requires the Hessian of the cost function, which will require calculations of the Hessians of the intensities at the individual detector pixels.
Calculating the Hessian matrix for each of the $L$\ locations $\{ \bs_l \}$ on the detector requires determining  $ L(N^2 + N)/2$\ quantities.
Even though the matrix elements in Eqs.~(\ref{eq: PhasePixelIntensityD2}) and~(\ref{eq: PhasePixelIntensityD2diag}) can be calculated with only a few FLOPs (once the field Jacobian has been calculated), nevertheless, this could be a daunting task under real-time constraints.  
Approximation of the Hessian of the intensity at a given pixel by a sparse matrix may be a useful strategy to speed real-time computation.

The sparsity pattern will depend on the pixel index $l$, but it can be pre-computed, since the particular values of the parameters $\{ c_k \}$\ only serve to rotate the field Jacobian values in the complex plane [see Eqs.~(\ref{eq: PhasePixelDetFieldD1}) and~(\ref{eq: PhasePixelIntensityD2})].
In order to gain a first look at the effective sparsity of one of these Hessian matrices, we used the PyWFS simulations described later to examine the Hessian matrix associated with the intensity at the central pixel in the upper-left pupil image (see Fig.~\ref{fig: PyramidImages}).
Let us denote that pixel with the index $l$.
The sparsity was evaluated on the matrix $<|\bH_l|>$, which is the absolute value of each matrix element averaged over 100 random realizations of the parameters $\{ c_k \}$.
In each of the 100 random realizations, all of the $N$\ values of $\{ c_k \}$ were given (independent) random values between $0$ and $2\pi$.
If $h_l^\mathrm{max}$\ represents the largest entry in  $<|\bH_l|>$, we found that about $0.51\%$\ of the matrix elements are greater than $0.01 h_l^\mathrm{max}$\ and $5.7\%$\ of them are greater than  $0.001 h_l^\mathrm{max}$.

Determining the utility of sparse approximations to these Hessians is beyond the scope of this article, as it must be done within the context of a detailed study of massively parallel implementations of the various estimation algorithms on a real-time system.
We only mention it because it may be useful in the future.

\begin{table}[t!]
\begin{center}
\begin{tabular}{|c | c | c | c | c| }
\hline
item & size  & $\times$ & $+$ & $\exp $ \\
\hline
$P\big( \be_k ; \, \bs_l \big) $ & $NL$ & p & p & p \\
\hline
$\frac{\partial}{\partial c_m} u_d(\bs_l; \bc)$ &  $NL$  & $2NL$ & 0 & $N$ \\
\hline
$u_d(\bs_l, \bc)$ & $L$  & $NL$ & $NL$ & 0 \\
\hline
$I_d(\bs_l, \bc)$ & $L$  & $L$ & 0 & 0 \\
\hline
$\frac{\partial}{\partial c_m} I_d(\bs_l; \bc)$ & $NL$ & $NL$ & 0  & 0 \\
\hline
$\frac{\partial^2}{\partial c_n \partial c_m} I_d(\bs_l; \bc)$ &  $N^2L/2$   & $ N^2L/2^\dagger$ & 0  & 0 \\
\hline 
\end{tabular}
\end{center}
\caption{\small 
Item sizes and approximate operation counts required for calculating various quantities needed for a single evaluation of the of Hessian of the intensity at every detector pixel, using the PhasePixel representation under the assumption of unit amplitude (i.e., $a_k = 1$).
There are $L$ pixels in the detector plane and $N$\ phases to be estimated, and typical numbers may be $L \approx 10^4$\ and $N \approx 10^3$.
The symbol "p" means "pre-computed," so that real-time constraints do not apply.
The operations counts for an item assume that computations for any items listed above it are re-used where possible.
For example, the second item in the table is $u_d(\bs_l, \bc)$, and the values given are for the entire set $\{ u_d(\bs_l, \bc) \}$.
In this case, the quantities given would be $L$\ times that required for evaluating it only at the point $\bs_l$.
This table gives arithmetic operations on complex numbers the same weight as real-valued numbers.
$^\dagger$Does not include the possible benefit of sparsity in the intensity Hessian discussed in Section~\ref{sec: discretization}.\ref{sec: sparsity}.}
\label{table: CompCount}
\end{table}

\section{Estimation Procedures}\label{sec: Estimation Procedures}

The goal of the inference procedure is to estimate the parameters $\bc$ from the intensity measurements.
The intensity values measured by the detector are noisy versions of the true intensities and can be modeled as:
\begin{equation}
z_l = I_d(\bs_l; \bc)  + \nu_l \, ,
\label{eq: y_l}
\end{equation}
where $l$ is the detector pixel index, $\{ z_l \}$ are the $L$\ measured intensity values and $\{\nu_l \}$ represents the (unknown) contribution of noise to the measurement, e.g., from photon counting (shot) noise and detector readout noise.  
It will be assumed that the $\{\nu_l \}$ are samples from zero-mean random process so that $E(\nu_l) = 0$, where $E$ is the expectation operator.

In this section, two estimation paradigms will be considered, linear least-squares and nonlinear least-squares.
Linear least-squares is representative of control algorithms that use a linear response matrix to estimate the wavefront.
Nonlinear least-squares can be more accurate than linear least-squares because the model $ I_d(\bs_l; \bc)$ is nonlinear in $\bc$.

\subsection{Preprocessing and Amplification of Nonlinearity Error}\label{sec: preprocessing}

\subsubsection{Preprocessing Options}\label{sec: options}

There are a number of options for pre-preprocessing the measured intensities before applying an estimation algorithm.
After this choice has been made, one can generate the vector of preprocessed measurements $\by$, where $\by$\ is used as input into a regression scheme.
Let $P$\ be the number of detector pixels within a single pupil image, so that geometrical optics would predict a total of $4P$ illuminated pixels.
Here, we will cover four preprocessing choices.
Two of these choices are slope preprocessing, which is common in the PyWFS community, and two are non-slope preprocessing.
Each preprocessing option is given a name:
\begin{enumerate}
\item{"FourImages" is simply the raw intensity values of the pixels within the four geometrical images of the pupil.
It is an example of non-slope preprocessing.
In this case, the $\by$ vector has $4P$ components. }
\item{"AllPixels" is the raw intensity values of all of the pixels on the detector, whether or not they are within one of the four geometrical images of the pupil, and it is also non-slope preprocessing.
In this case, $\by$ has $L$ components where $L$ is the total number of detector pixels.}
\item{"NomalizedSlope" is the most common in the AO community, in which the 4 pupil images are reduced to two "slope" images.\cite{Verinaud2004PyWFS} 
These images are produced by adding and subtracting pairs of images, and dividing by the mean image.  This results in a $\by$ having $2P$ components.}
\item{"UnnormalizedSlope" is the same NormalizedSlope, except there is no division by the mean of the four pupil images.  Here, $\by$ also has $2P$ components.}
\end{enumerate}

Thus, with FourImages and AllPixels, the components of $\by$ have a very simple relationship to the values of detector intensities in \eqref{eq: y_l}, namely, 
\begin{equation}
\overset{\mathrm{FourImages , \, AllPixels}}{
y_l}
  = z_l \, .
\label{eq: AllPixels}
\end{equation}
Under the UnnormalizedSlope preprocessing, $\by$ represents two slope images, so $\by$ can be split into a vertical slope image $\by^\mathrm{v}$\ and a horizontal slope image $\by^\mathrm{h}$, in which the elements are given by:
\begin{align}
\overset{\mathrm{UnnormalizedSlope}}{ 
y_l^\mathrm{v}} \; = \; & z_l^{++} + z_l^{-+} -  z_l^{--} - z_l^{+-}  \nonumber \\
\overset{\mathrm{UnnormalizedSlope}}{
y_l^\mathrm{h}} \; = \; & z_l^{++} + z_l^{+-} -  z_l^{-+} - z_l^{--} \, , 
\label{eq: unnormalized slope}
\end{align}
where $z_l^{++}$\ is the intensity measured in the $l$\underline{th} pixel in the upper-right pupil image, and $z_l^{-+}$,  $z_l^{--}$, $z_l^{+-}$ are the corresponding values in the  upper-left, lower-left and lower-right pupil images, respectively.
NormalizedSlope preprocessing is similar, except that the horizontal slopes are divided by the mean of the four pupil images, so:
\begin{align} 
\overset{\mathrm{NormalizedSlope}}{
y_l^\mathrm{v}} \; = \; & 4 \frac{z_l^{++} + z_l^{-+} -  z_l^{--} - z_l^{+-}}
{z_l^{++} + z_l^{-+} +  z_l^{--} + z_l^{+-}}  \nonumber \\
\overset{\mathrm{NormalizedSlope}}{
y_l^\mathrm{h}} \; = \; & 4 \frac{z_l^{++} + z_l^{+-} -  z_l^{-+} - z_l^{--}}
{z_l^{++} + z_l^{-+} +  z_l^{--} + z_l^{+-}}  \, .
\label{eq: normalized slope}
\end{align}
Note that if the noise in the $z_l$ values is modeled by a zero-mean random process $\nu_l$\ as in \eqref{eq: y_l}, then the division in \eqref{eq: normalized slope} implies that NormalizedSlope values will have noise statistics that are not given by a zero-mean random process, whereas the other three representations have noise statistics that are straightforward (given the processes $\{ \nu_l \}$).

\begin{figure}[t]
\includegraphics[height=60mm]{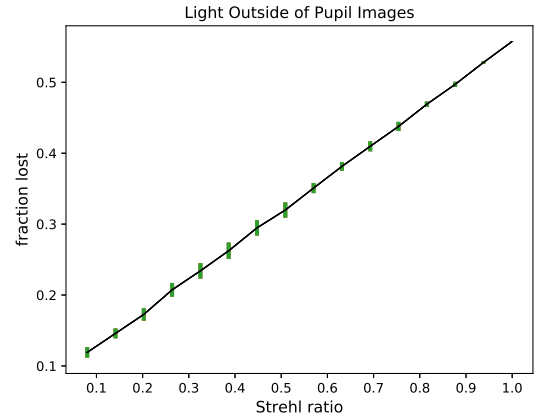}
\caption{\small  The fraction of the intensity falling outside of the geometrical images of the four pupils as a function of Strehl ratio of the input beam.  Each data point is the mean result of 60 trials with random phases.
The error bars indicate the standard deviation of the 60 results.  This figure indicates that, for the PyWFS with the parameters simulated here, about 56\% of the light falls outside of the pupil images as the Strehl ratio approaches unity.
}
\label{fig: light loss}
\end{figure}

One fact that should be appreciated is that a signficant fraction of the light can fall outside of the 4 four geometrical pupil images.
The amount of light lost by only collecting the light within the four geometrical pupil images increases with Strehl ratio.
This effect is shown in Fig.~\ref{fig: light loss} for the simulation experiments described in detail later.
The figure shows that when the Strehl ratio is 0.1 only about 10\% of the light is lost, but that fraction increases to about 56\% at a Strehl ratio of 1.0.

\begin{figure}[t]
\includegraphics[height=60mm]{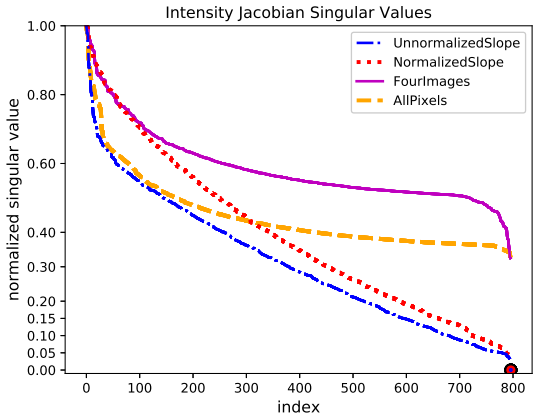}
\caption{\small  Normalized "spectrum" of singular values of the Jacobian (evaluated for a flat wavefront) under the PhasePixel representation of the wavefront for slope and non-slope preprocessing options.
In all cases, the final singular value is zero, corresponding to the unobservable piston term.
While non-slope preprocessing options have no singular values less than about 0.4 (other than the piston term), the slope preprocessing options have more than one-half of their singular values below 0.4, leading to more amplification of the nonlinearity error in the estimated phases, as can be seen in Fig.~\ref{fig: nonlinear error}.}
\label{fig: singular values}
\end{figure}

\begin{figure}[t]
\begin{tabular}{l}
\includegraphics[height=60mm]{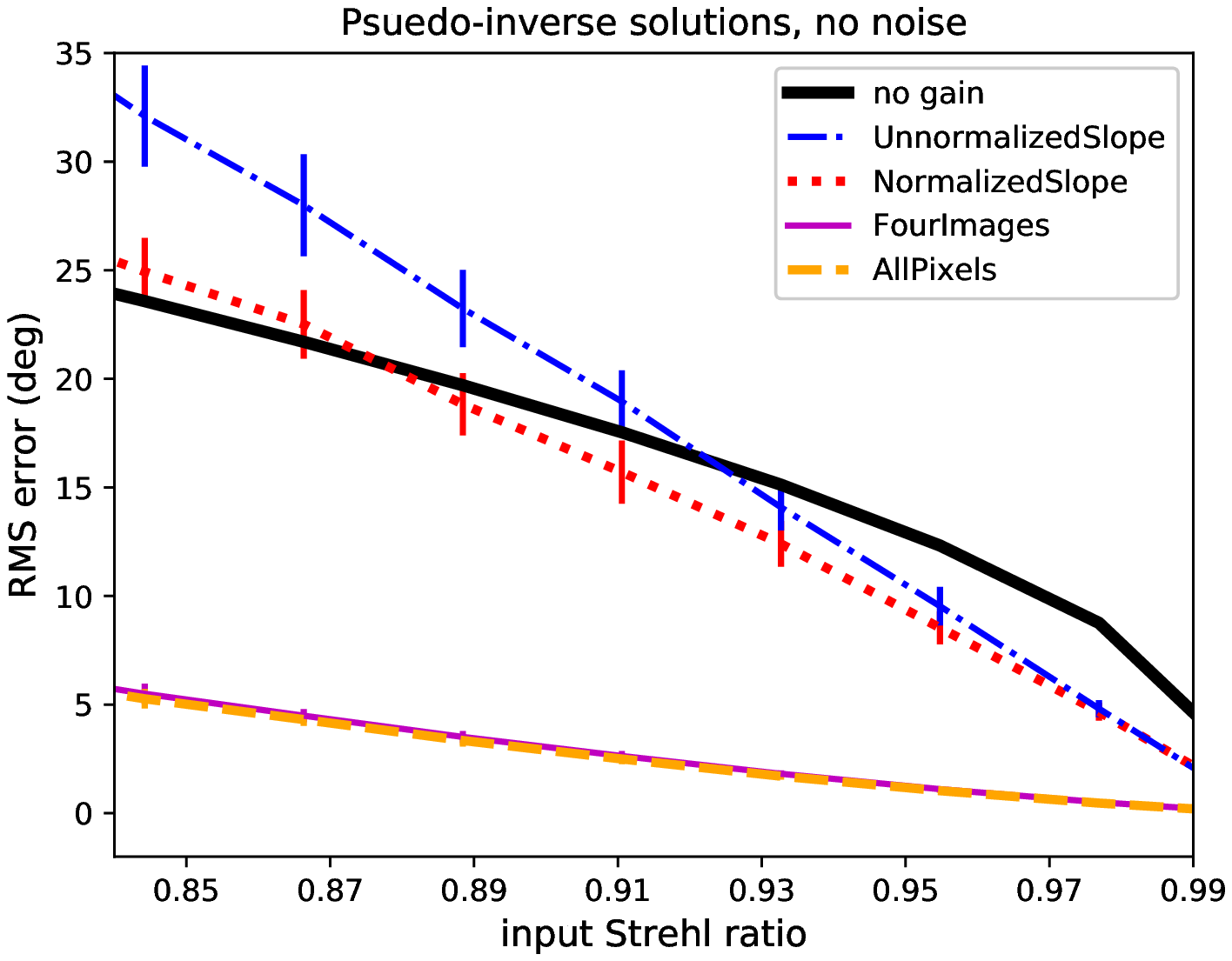} \\
\includegraphics[height=60mm]{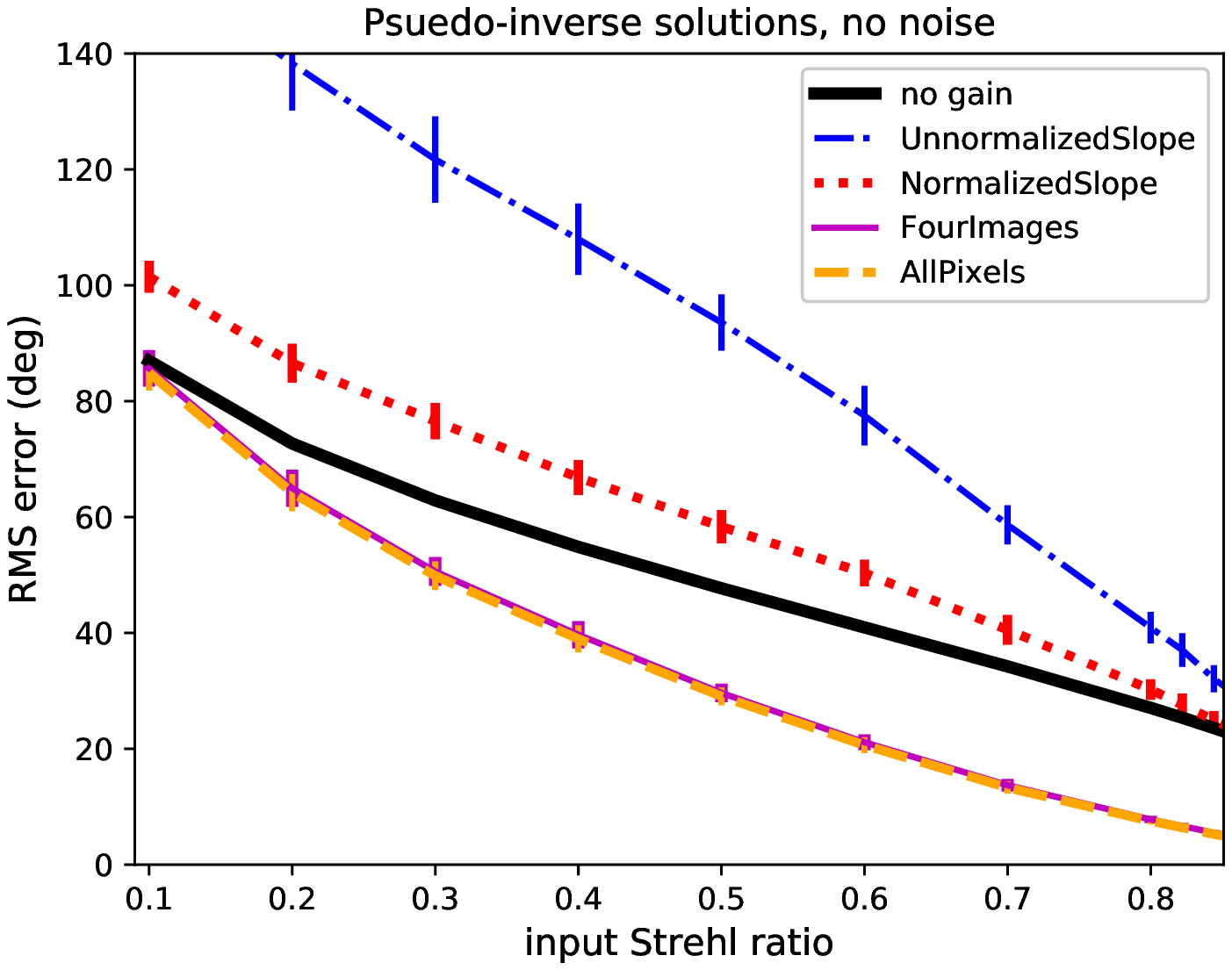}
\end{tabular}
\caption{\small  RMS error of the pseudo-inverse phase estimate for slope (NormalizedSlope, UnnormalizedSlope) and non-slope (FourImages, AllPixels) preprocessing options, as a function of the Strehl ratio of the input beam (note that the pseudo-inverse is one type of linear estimator).
As no noise was included, this error is entirely due to the nonlinearity error.
Each data point is the mean RMS error over 60 trials, and the error bars indicate the standard deviation of these RMS errors over the 60 trials.
This figure shows that the slope preprocessing only provides gain when the input Strehl ratio is greater than about $0.9$, while non-slope preprocessing provides gain down to a Strehl ratio of about 0.1.
This is due to the fact that slope preprocessing leads to more amplification of the nonlinearity error, as explained in the text.
Note that the two non-slope curves are nearly overlapping and indistinguishable in the figure.
 \emph{top:} Strehl ratio $\geq 0.85$.   \emph{bottom}: Strehl ratio $< 0.85$. 
}
\label{fig: nonlinear error}
\end{figure}

\subsubsection{Information Loss in Slope Sensing}\label{sec: slope sensing}

Besides discarding some of the light, as shown in Fig.~\ref{fig: light loss}, there is a more subtle, yet far more detrimental problem associated with slope sensing with the non-modulated PyWFS.
This is the error caused by amplification of the nonlinearity by the inversion used to obtain the phases.
Developing these arguments will require a bit of detailed treatment of the error incurred by first order Taylor expansions (nonlinearity error) and its consequences for linear regression.
Consider the following generic nonlinear regression model:
\begin{equation}
\by = \bA(\bc) + \bnu\, , 
\label{eq: general nonlinear model}
\end{equation}
where $\by$\ has $L$ components (the number measurements), $\bc$ is a vector of unknown values with $N$\ components that we wish to determine, $\bA(\bc)$\ is a vector of $L$\ nonlinear functions of $\bc$, and $\bnu$\ is a vector of zero-mean random processes representing noise.
Sometimes $\bA(\bc)$\ is called the measurement function, since it relates $\bc$\ to the measured values.
The use of $\by$, $\bc$, $N$ and $L$\  is consistent with our previous uses of these symbols above, but for this section, they are intended only to describe the generic nonlinear regression model shown in \eqref{eq: general nonlinear model}.
The Taylor expansion of $\bA(\bc)$\ about the point $\bc_0$\ can be written in terms of $\bA$'s Jacobian $L \times N$\ matrix $\bG$, so that the measurement function can be rewritten as:
\begin{equation}
\bA(\bc) = \bA(\bc_0) +  \bG[\bc - \bc_0] + \bB(\bc - \bc_0) \, ,
\label{eq: A linearization}
\end{equation} 
where $\bG \equiv \partial \bA / \partial \bc |_{\bc_0}$ is the Jacobian matrix, and the new vector of functions $\bB$\ (with $L$ components) accounts for the error incurred by the linearization, such that $\bB(0) = 0$.
The vector of functions $\bB$\ will be called the "nonlinearity error."
To economize, let us make the following change in notation: $\bc - \bc_0 \rightarrow \bc$, $\by - \bA(\bc_0 )\rightarrow \by$.
Then \eqref{eq: general nonlinear model} becomes:
\begin{equation}
\by  =   \bG \bc + \bB( \bc ) + \bnu \, ,
\label{eq: linearization}
\end{equation}
If we ignore the nonlinearity error $\bB(\bc)$, which is necessary for linear regression, it can be shown that the least-squares estimate of minimum norm is provided by the pseudo-inverse of the matrix $\bG$, written as $\bG^+$, via the equation:
\begin{equation}
\hat{\bc} = \bG^+ \by  \, .
\label{eq: pinv estimate}
\end{equation}
In order find $\bG^+$, first perform the singular value decomposition (SVD) of $\bG$ as $\bG = \bU \bSigma \bV^\rT$ (the superscript $^\rT$\ indicates matrix transposition), where $\bU$\ and $\bV$\ are unitary matrices and $\bSigma$\ is a diagonal matrix of singular values $\{ \sigma_k \}, \, 0 \leq \sigma_k $.
The set of singular values $\{ \sigma_l\}$, ordered from largest to smallest, is also known as the \emph{spectrum} of the matrix or the spectrum of singular values.
Then $\bG^+ \equiv \bV \bSigma^+ \bU^\rT$, where $\bSigma^+$\ is a diagonal matrix of the inverses of all of the non-zero singular values, i.e.,  $1/\sigma_k  \, \mathrm{for \; all}\, \sigma_k > 0 $.
Where the singular values are zero (or deemed to be numerically small) the corresponding entries in the diagonal of $\bSigma^+$ are assigned values of zero.\cite{Moon&Stirling}

In order appreciate how the inversion in \eqref{eq: pinv estimate} can amplify the noise $\bnu$ and the nonlinearity error $\bB(\bc)$, substitute \eqref{eq: linearization} into \eqref{eq: pinv estimate}:
\begin{align}
\hat{\bc} & =  \bG^+ \bG \bc +  \bG^+\big[  \bB( \bc ) + \bnu \big]  \nonumber \\
& =    \bV \bSigma^+ \bSigma \bV^\rT  \bc + 
\bV \bSigma^+ \bU^\rT \big[  \bB( \bc ) + \bnu \big] 
\label{eq: pinv estimate noise}
\end{align}
where the unitary property of $\bU$ (i.e., $\bU^\rT \bU = \mathbb{1} $, where $\mathbb{1}$ is the identity matrix) has been utilized.
The matrix $\bSigma^+ \bSigma $ is a diagonal matrix with values of unity where $\sigma_l > 0$ and 0 where $\sigma_l = 0$ (or small enough numerically to be considered zero).
In the first term of \eqref{eq: pinv estimate noise}, the components of $\bc$ receive no amplification since $\bV$ is unitary.
It is easy to show that $| \bV \bSigma^+ \bSigma \bV^\rT  \bc | \leq |\bc|$.
Further, if all of the singular values are greater than zero (i.e., $ \sigma_l > 0 \; \mathrm{for \, all} \, l$), then $\bV \bSigma^+ \bSigma \bV^\rT \bc = \bc $.
On the other hand, the small singular values in $\bSigma^+$\ in the second term of \eqref{eq: pinv estimate noise} will amplify both the nonlinearity error $\bB( \bc ) $ and the noise $\bnu$.
Thus, small singular values are detrimental to the accuracy of the inversion even in the case in which there is no noise (i.e., $\bnu = 0$).
One popular method of mitigating this noise amplification, called SVD truncation, is to replace $1/\sigma_k$\ with 0 on the diagonal of $\bSigma^+$\ whenever $\sigma_k$ is below some threshold.
When the SVD is truncated, the associated modes are forced to 0.
[The modes under discussion here are singular vectors (i.e., the columns of the matrix $\bV$), not the $\psi$ functions in \eqref{eq: ModalPhase}.]

NormalizedSlope, UnnormalizedSlope, FourImages and AllPixels (see Sec.~\ref{sec: Estimation Procedures}\ref{sec: options}) each correspond to a different version of the measurement function $\bA(\bc)$, according to Eqs.~(\ref{eq: AllPixels}) through (\ref{eq: normalized slope}), each of which relates to the computational model of the detector intensity via \eqref{eq: y_l}.
Then, the calculation of the Jacobian matrix $\bG$\ for each preprocessing option can be calculated using \eqref{eq: PhasePixelIntensityD1}.
Fig.~\ref{fig: singular values} shows the resulting singular values of $\bG$, where it can be seen that the singular values obtained using slope preprocessing (NormalizedSlope and UnnormalizedSlope) decline more steeply than they do for non-slope preprocessing (FourImages and AllPixels).
For each case (NormalizedSlope, UnnormalizedSlope, FourImages and AllPixels), the relevant Jacobian was evaluated for a flat wavefront, i.e., $\bc_0 = 0$, and the singular values were calculated with a standard SVD algorithm.
Each curve in Fig.~\ref{fig: singular values} is normalized by its first singular value, which is the largest one.
In all cases, the final singular value is 0, corresponding to the piston term, which cannot be determined from a WFS.
(The simulations showed that the final column of the $\bV$ matrix is a vector in which all values are the same, confirming that the final singular value does indeed correspond to the piston term.) 

The more rapid decline in singular values for the slope preprocessing has dramatic consequences for estimation of the wavefront.
Fig.~\ref{fig: nonlinear error} shows RMS phase error obtained by applying the pseudo-inverse formula [\eqref{eq: pinv estimate}], as a function of the input Strehl ratio of the input beam.
Each data point in Fig.~\ref{fig: nonlinear error} is a result of 60 trials in which the 797 phase values were sampled independently from a normal distribution with a standard deviation given by by $\sqrt{-\ln S }$ (Ruze formula), where $S$ is the Strehl ratio of the input beam.
The error bars (at least those large enough to be visible) indicate the standard deviation of the RMS error over the 60 trials.
The piston term (i.e., the mean) of these randomly generated phase values was subtracted in each trial, so no error was caused by the final zero on the diagonal of $\bSigma^+$, which forces the piston term of the estimate to be zero.
All of the trials used to create this figure had no noise, so $ \bnu = 0$.
Therefore, all of the errors in the estimates $\hat{\bc}$ shown in the figure were caused by the nonlinearity error $\bB(\bc)$.
The figure shows that the inversions corresponding to FourImages and AllPixels far outperform the inversions corresponding to NormalizedSlope and UnnormalizedSlope, with UnnormalizedSlope being the slightly better of the two.

The black curve labeled "no gain" in Fig.~\ref{fig: nonlinear error} corresponds to an RMS error equal to $\sqrt{-\ln S }$.
Whenever an estimator is below this curve it is providing some information about the wavefront, but when an estimator is above this curve, it is performing more poorly than simply using $\hat{\bc} = 0$ as the estimator.
The figure shows that NormalizedSlope and UnnormalizedSlope only provide gain for Strehl ratios greater than about 0.9,  and actually provide negative gain for lower Strehl ratios.
In contrast, the estimators corresponding to FourImages and AllPixels provide gain for all Strehl ratios in the figure, down to a Strehl ratio of about 0.1.
In order for the slope preprocessing to provide gain at lower Strehl ratios, it would be necessary to either apply SVD truncation as explained above, and/or use a ModalPhase representation with a number of modes that is small enough to avoid problematic singular values.
Both of these approaches effectively trade numerical stability for decreasing the spatial resolution of the estimated wavefront.

Thus, this analysis indicates that the traditional slope preprocessing is likely to prove suboptimal for the non-modulated PyWFS and it will not be considered further in the remainder of this article.
In addition, from this point forward, we will only consider the AllPixels preprocessing option, since the FourImages discards some of the light at high Strehl ratio, as Fig.~\ref{fig: light loss} shows.

\subsection{Linearized Least-Squares with Regularization}

The pseudo-inverse was introduced above in order to show how inversion can amplify noise and error due to nonlinearity.  
While one could perform least-squares estimation from noisy data by employing a truncated pseudo-inverse, this method does not generalize to nonlinear estimation.
Instead, here we introduce a regularization term that mitigates the effects of small singular values in the Jacobian matrix.
The same term can be included in nonlinear estimation procedures, as we demonstrate later.

Going forward under the AllPixels preprocessing for the reasons given above, the vector of intensity measurements $\by$\ has components defined by Eqs.~(\ref{eq: y_l}) and (\ref{eq: AllPixels}).
It will economize the notation to redefine $\bc$\ and $\by$ via: $\bc - \bc_0 \rightarrow \bc$, $\by - I_d(\bs; \bc_0) \rightarrow \by$,  where $I_d(\bs; \bc_0)$ is all $L$ values of $I_d(\bs_l; \bc_0)$ arranged into a $L \times 1$ column vector.
The unknown parameters $\bc$ are related to the measurements via \eqref{eq: linearization}.
Here, the symbols are defined specifically for the PyWFS estimation problem, not in generic terms as they were when \eqref{eq: linearization} was first presented.
The linearized model can be expressed in terms of the $L \times N $\ Jacobian matrix $\bG$:
\begin{equation}
\bG \equiv \frac{\partial}{\partial \bc} I_d(\bs_l; \bc) \bigg|_{\bc_0} \, ,
\label{eq: Hdef}
\end{equation}
which is the Jacobian evaluated at $\bc_0$.
In practice, $\bc_0$\ would correspond to a flat wavefront (zero phase) or take into account known static aberration.
The Jacobian can be calculated from Eq.~(\ref{eq: PhasePixelIntensityD1}) for the PhasePixel representation.The linearized least-squares solution is a minimizer of the (scalar) cost function:
\begin{equation}
\mathcal{C}_\mathrm{ l}(\bc) = \frac{1}{2}[\bG \bc - \by]^\rT \bW [\bG \bc -\by] +  \frac{1}{2}\alpha \bc^\rT \bR \bc \, ,
\label{eq: CostLLS}
\end{equation}
where $\bW$\ is a matrix of the measurement weights ($\bW$ is diagonal for uncorrelated noise), $\bR$\ is a regularization matrix and $\alpha$\ ($ \geq 0$) is a regularization parameter that controls the strength of the regularization.
Taking the measurement weight $\bW$\ to be the inverse of the measurement noise covariance matrix corresponds to the maximum-likelihood cost function for Gaussian noise statistics.
The minimizer of Eq.~(\ref{eq: CostLLS}) is easily shown to be:
\begin{equation}
\hat{\bc} = [\bG^\rT \bW \bG + \alpha \bR]^{-1} \bG^\rT \bW \by
\label{eq: estimateLLS}
\end{equation}
Then, the linear least-squares estimate is $\bc_1 = \hat{\bc} + \bc_0$.
A proper choice of $\bR$ and non-zero choice of $\alpha$\ ensures invertibility of the matrix $[\bG^\rT \bW \bG + \alpha \bR]$ in Eq.~(\ref{eq: CostLLS}); also see comments in Ref.~[\citenum{KorkiPyWFSreconNonlin2007}].
Once the quantities $\bc_0$, $\alpha$, $\bR$ and $\bW$ have been chosen, the $N \times L$ matrix $[\bG^\rT \bW \bG + \alpha \bR]^{-1} \bG^\rT \bW $\ can be pre-computed, so that calculating the linear least-squares estimate only involves a single matrix-vector multiplication.

Common choices for $\bR$ include the $N \times N$\ identity matrix and finite differencing operators implemented with values of $o(1)$\ (where $o(1)$\ means "of order unity") on the appropriate diagonals.
For example, if $\bD$\ is one such matrix, then $\bR = \bD^\rT \bD$.
Matrices of this type have a largest singular value on the order of unity, so that a baseline value for $\alpha$ can be calculated by $\alpha_0 = \by_0^\rT \bW \by_0$\ where $\by_0$\ is the vector of synthetic intensity values associated with a flat wavefront (i.e., $\by_0 = \{ I_d(\bs_l; \bc_0) \}$ in the PhasePixel representation).
Then, $\alpha = \alpha_0$\ would correspond to a solution dominated by regularization, and we would expect optimal values of $\alpha$ to correspond to $\alpha = q \alpha_0$, where $ 0 \leq q < 1$.

\subsection{Nonlinear Least-Squares}\label{sec: NLS}

The purpose of the nonlinear least-squares is to account for the fact that the measurement function $ I_d(\bs_l; \bc) $\ is nonlinear in the parameters $\bc$.
The nonlinear cost function corresponding to Eq.~(\ref{eq: CostLLS}) is:
\begin{equation}
\mathcal{C}_\mathrm{nl}(\bc) = \frac{1}{2} \big[ I_d(\bs; \bc) - \by \big]^\rT \bW \big[I_d(\bs; \bc) - \by \big] +  \frac{1}{2}\alpha \bc^\rT \bR \bc \, .
\label{eq: CostNLS}
\end{equation}
\eqref{eq: CostNLS} does not have $\by$\ defined with $I_d(\bs; \, \bc_0)$ subtracted,  nor is $\bc$ defined with $\bc_0$\ subtracted, as they were in \eqref{eq: CostLLS}, since \eqref{eq: CostNLS} is not a linearization.
The discussion below will require the gradient and Hessian of the cost function:
\begin{align}
\frac{\partial}{\partial c_m} \mathcal{C}_\mathrm{nl}(\bc)  = & \sum_{l=0}^{L-1} \left\{  W_{ll} \big[I_d(\bs_l; \bc) - \by \big] 
\frac{\partial}{\partial c_m} I_d(\bs_l; \bc)
+ \alpha R_{ml} c_l \right\} 
\label{eq: CostNLD1} \\ 
\frac{\partial^2}{\partial c_n \partial c_m} \mathcal{C}_\mathrm{nl}(\bc)  = & \sum_{l=0}^{L-1} \bigg\{ 
W_{ll} \frac{\partial}{\partial c_n} I_d(\bs_l; \bc) \frac{\partial}{\partial c_m} I_d(\bs_l; \bc) \; + \nonumber \\
 & \: \: \: \: \: \: \: \: \: \: \:
W_{ll} \big(I_d(\bs_l; \bc) - \by \big)  \frac{\partial^2}{\partial c_n \partial c_m} I_d(\bs_l; \bc) + \alpha R_{mn} \bigg\} \, ,
\label{eq: CostNLD2}
\end{align}
where $W_{ll}$\ is the $l$\underline{th} element on the diagonal of $\bW$\ and $R_{ml}$\ is an element of the regularization matrix $\bR$.
Note that \eqref{eq: CostNLD1} and \eqref{eq: CostNLD2} assume that the weight matrix $\bW$\ is diagonal.
The needed derivatives of the intensity are shown in Eqs.~(\ref{eq: PhasePixelIntensityD1}) through (\ref{eq: PhasePixelIntensityD2diag}) for the PhasePixel representation.
Since each element of the $N \times N$\ Hessian of the cost function requires a sum over $L$\ pixels, calculation of this Hessian requires the determination of $(N^2 + N)L/2$ quantities.
While the Hessian of the cost function is not itself sparse, the intensity Hessians in the second term of Eq.~(\ref{eq: CostNLD2}) may have useful sparse approximations, as already discussed in Sec.~\ref{sec: discretization}\ref{sec: sparsity}.

\eqref{eq: CostNLS} represents an unconstrained nonlinear optimization problem, in which one attempts to find the value of the parameter vector $\bc$\ that minimizes $\mathcal{C}_\mathrm{nl}(\bc)$.
Unconstrained nonlinear optimization has two goals that often oppose each other: finding a minimum with minimal computational cost and avoiding local minima.
In some optimization problems (e.g., nonsingular quadratic cost functions), there is only one local minimum that is the global minimum.
Otherwise, avoiding local minima requires global minimization methods, such as simulated annealing, that involve stochastic searching of the parameter space.
These methods are very slow and are not suited for the needs of control systems or large batch estimation.
Local optimization methods, on the other hand, do not attempt to find a global minimum.
Instead they find the local minimum corresponding to the initial guess; one could say "they just go down the hill" \cite{BertsekasNLP}.
As nonlinear optimization is always iterative, it requires an initial guess.
One choice for this guess is the linear least-squares solution $\bc_1$, given above, and that is what has been used for the numerical examples shown later.

In terms of iteration counts, the most efficient local optimization method is often Newton's method, and it almost always converges in many fewer iterations than gradient descent.
This is because Newton's method is based on successively approximating the cost function with quadratic functions, as opposed to gradient descent, which does not account for the local curvature of the cost function.
Newton's method requires the gradient and Hessian of the cost function, shown here in Eqs.~(\ref{eq: CostNLD1}) and~(\ref{eq: CostNLD2}) for the PhasePixel representation.
In its purest form, the Hessian matrix needs to be inverted, but in practice this is avoided by solving a system of equations iteratively with the conjugate gradient algorithm.
Thus, the cost of a single Newton iteration is often dominated by calculating the Hessian  and performing the multiple iterations of the conjugate gradient algorithm.
Calculating the Hessian matrix is sometimes prohibitive and this has given rise to a variety of  "quasi-Newton" methods that only require the gradient of the cost function, avoiding explicit calculation of the Hessian.
One widely popular quasi-Newton method is the  Broyden–Fletcher–Goldfarb–Shanno (BFGS) algorithm, and that is the algorithm we discuss here and implement in the simulation experiments below.
The BFGS algorithm updates an approximation to the inverse of the Hessian of the cost function with every iteration, so it almost always takes multiple BFGS iterations to achieve the minimization attained by a single Newton iteration.
The cost of a single BFGS iteration has three primary contributors: a line search requiring multiple function and gradient evaluations, and a matrix-matrix multiplication required to update the approximation to the inverse Hessian.\cite{BertsekasNLP}
Whether an implementation of Newton's method or a quasi-Newton method would be more efficient for real-time nonlinear estimation for a PyWFS depends on the details of how the parallelization is implemented and how much sparsity can reduce the cost of calculating an approximate Hessian (see Sec.~\ref{sec: discretization}\ref{sec: sparsity}).
A detailed comparison of Newton and quasi-Newton methods for nonlinear PyWFS estimation under different parallelization schemes is beyond the scope of this article.
Instead, this article argues that the nonlinear cost function, its gradient and its Hessian can be implemented efficiently and in parallel, thus enabling, in principle, both Newton and quasi-Newton methods.

\subsection{Enforcing Zero-Mean Phase}\label{sec: zero-mean}

The WFS has no response to the piston term, causing a zero singular value in the Jacobian matrix of the measurement function (see Fig.~\ref{fig: singular values}).
Depending which representation of the wavefront is being used, there are several ways to force the piston term to be zero and resolve the issue of the zero singular value.
If one is using ModalPhase, simply remove the piston term from the expansion in \eqref{eq: ModalPhase}.
In this way, the singular value of 0 never arises.
This approach assumes the remaining modes have no piston component.
In the PhasePixel representation when using linearized least-squares, augmenting the Jacobian $\bG$\ with a row of $1/N$ and appending $\by$  with 0 (a weight value, $\beta$, should also be appended to the diagonal of $\bW$) explicitly forces the mean of the estimated phases $\hat{\bc}$ to be 0.
This makes the singular value corresponding to the piston term finite.
When using nonlinear least-squares with the PhasePixel representation, adding a another term, $\beta (\sum_n c_n  )^2/N^2$ (where $\beta$ is a weight), to the cost function in Eq.~(\ref{eq: CostNLS}) achieves the same end.
In practice, the weight value $\beta$\ should be large enough that the mean phase is quite close to zero in the final solution, but not large enough to cause numerical difficulties due to finite floating-point precision.

\subsection{Application to Time-Series}

So far, we only addressed the problem of estimating a single wavefront, not a time-series of wavefronts.
The problems should be treated differently since the wavefront at $t$\ is highly correlated with the wavefront at time $t + \Delta t$, where $\Delta t$\ is the time-step of the AO loop, typically on the order of 1 ms.
Thus, an optimal estimation method would take advantage of this information.
The typical solution in linear estimation is Kalman filtering, and there are standard extensions to treat nonlinearities.
Kalman filtering can become computationally expensive due a certain matrix inversion, and there is a large literature devoted to dealing with this difficulty.\cite{Anderson&Moore}
One way to account for the temporal correlation in the wavefront within the framework already presented is to add the following term to the cost function when calculating the solution for $t + \Delta t$:
\begin{equation}
\Delta \mathcal{C}(\bc) =  \frac{1}{2} \gamma [\bc - \hat{\bc}_t]^\rT \bC [\bc - \hat{\bc}_t] \, ,
\label{eq: temporal_pentalty}
\end{equation}
where $\hat{\bc}_t$\ is the estimate from the previous time-step, $\gamma$\ is a weight, and $\bC$\ is an $N \times N$ matrix.
Note that if $\bC$ is taken to be the identity matrix, then the penalty term in \eqref{eq: temporal_pentalty} has a form that is similar to the regularization penalty in Eqs.~(\ref{eq: CostLLS}) and~(\ref{eq: CostNLS}), and it can be used to mitigate the effects of singularity or near-singularity in the intensity Jacobian [\eqref{eq: PhasePixelIntensityD1}].

Another important aspect of applying the linear and nonlinear least-squares algorithms to time-series concerns the choice of the starting guess.
The estimate at the previous time-step, $\hat{\bc}_t$, is an obvious choice; another choice would be $\mu \hat{\bc}_t$, where $0 \leq \mu < 1$\ in order to improve stability.

\
\section{Simulation Experiments}

A numerical simulation of the PyWFS with geometrical parameters similar to the PyWFS on the SCExAO/Subaru generated the results shown in this article.\cite{SCExAO_PASP15}
The prism is taken to be a square pyramid with each face making an angle to the horizontal of $3.73^\circ$ and an index of refraction of 1.452 at the operating wavelength of $0.85 \; \mu$m. 
The telescope optics reduce the $D \approx 8 \;$m diameter beam down to a diameter of $d = 7.2$ mm, and a lens with focal length $f = 40d$\ focuses the beam onto the tip of the pyramid.

Numerically, the beam in the entrance pupil was sampled with 33 pixels across, resulting in 797 pixels in the circular pupil (no "spiders" or other obscurations were included).
In order to simulate a wavefront under the PhasePixel representation, each of the $N\,= \, 797$ entrance pupil pixels had a phase $ c_k$\ (thereby creating the $\bc$ vector) assigned.  The 797 amplitude values, i.e., $\{ a_k \}$ in \eqref{PhasePixel}, were taken to be unity.
We did not consider scintillation effects.
In order to perform the Fourier transforms in Eq.~(\ref{eq: PyramidOp}), these 797 pixels (forming a disk) were embedded into the center of a $1024 \times 1024$ array of zeros.
To calculate the field at the focal plane (at the tip of the pyramid), a fast Fourier transform (FFT) was applied to the field values, resulting  in a $1024 \times 1024$ array that represents the field values in the focal plane.
Then, these $1024 \times 1024$ field values were multiplied by the phase ramp induced by the pyramid faces.
Finally, the four pupil images were formed by taking another FFT, resulting in a $1024 \times 1024$ array of field values in the detector plane, as per \eqref{eq: SecondFT}.
The "entire detector" was taken to be  the $125 \times 125 = 15625$ pixels in the center of the  $1024 \times 1024$\ array that represents the detector plane, and the AllPixels preprocessing refers to this set of pixels as input into the regression calculations.

We made the choice to create the phases from a white noise process in lieu of a turbulence power-law, since a PyWFS without modulation is likely to see a beam that has already passed through a first-stage AO system, as explained in the introduction.
A first-stage AO system would remove much of the low-order aberration, so, white noise seemed a reasonable choice that avoids the rigors of end-to-end simulation.
To simulate wavefronts corresponding to a given Strehl ratio $S$, the 797 "truth" values of $\bc$ were drawn independently from a normal distribution with a standard deviation $\sigma = \sqrt{-\ln S }$ (the Ruze formula).
By taking the phases in each pupil plane pixel to be statistically independent, we implicitly assume that the Fried parameter $r_0$ roughly corresponds to the length of an entrance pupil pixel (scaled up to the telescope diameter), in this case: $r_0 \approx D/33 \approx 24$\ cm. 
(In fact, since the RMS variation of the phase over a circle of diameter $r_0$ is about one radian, this is an under-estimate of the implied $r_0$.)

The regression tests used to evaluate the linearized and nonlinear estimators included noise from two sources: Poisson counting noise (shot noise) and detector readout noise.
The shot noise was approximated by a normal distribution with the standard deviation set to $\sqrt{\mathrm{intensity}}$ (which was known from the noise-free simulation).
The readout noise was modeled by a normal distribution with the standard deviation set equal to the detector noise level.
The noise from both of these sources was sampled independently for each detector pixel.

These simulations were not optimized for speed and no computations were implemented in parallel.
In this context, calculating the Hessian of the cost function made Newton's algorithm orders of magnitude slower than the BFGS algorithm, as implemented in the "optimize.minimize" module of SciPy (see \emph{http://www.SciPy.org}).
The BFGS algorithm was set to exit when the norm of the cost function gradient (normalized) dropped below $10^{-7}$, which usually required roughly 100 function and gradient evaluations.
Similar results were achieved by 10 or fewer Newton iterations, each of which requires one Hessian evaluation and one gradient evaluation.

\subsection{Results: Interference Effects}

The previously cited work by Korkiakoski et al. \cite{KorkiPyWFSreconNonlin2007} neglects interference effects when calculating the four pupil images.
In the Fourier optics model presented here, the interference is treated automatically since all four beams are propagated simultaneously.
In order to gain some understanding of the error due to neglecting interference between the two beams, we performed two simulations:
In the first simulation, we utilized the full Fourier optics model, thereby creating $I_\mathrm{t}$ (the "true" image).
In the second simulation, we neglected interference, thereby creating $I_\mathrm{a}$ (the "approximate" image).
We produced the interference-free simulation by blocking transmission in 3 of the 4 pyramid faces and recording the intensity (which only had a single pupil image).
One such image was created for each of the 4 pyramid surfaces, each in a different quadrant of the detector. 
Then, the 4 images were added to create the final, interference-free image $I_\mathrm{a}$.
For this example, the Strehl ratio of the input beam was 0.3, and no noise was included.

The resulting images $I_\mathrm{t}$ and $I_\mathrm{a}$ were visually similar, but the mean of the ratio $\eta \equiv |I_\mathrm{t} - I_\mathrm{a}| / I_\mathrm{t} $\ inside the pupil images was about 0.37.
The results of this calculation are displayed in Fig.~\ref{fig: PyramidImages}.
The upper panel of Fig.~\ref{fig: PyramidImages} shows $\sqrt{I_\mathrm{t}}$ (normalized intensity units), and the lower panel
displays the image of the log-ratio ($\log_{10} \eta $).
This result suggests that interference effects can be important and should not be neglected without justification.

\subsection{Results: Regression Tests}

The regressions were designed to compare the accuracy of the linear least-squares estimate of the phase of the field in the entrance pupil to the estimate made by nonlinear least-squares.
The parameters varied for the different simulations runs were: the Strehl ratio of the input beam, the total number of photons entering the PyWFS, and the detector readout noise.
The Strehl ratios for the input beam values were:  $0.05, \, 0.1, \, 0.2, \, 0.3, \, 0.4, \, 05, \, 0.6, \, 0.7, \, 0.8, \, 0.9$, the total photon counts were $10^4, \, 10^5, \, 10^7$, and the detector readout noise standard deviations $\sigma_\mathrm{r}$ were 0 and 4 counts/pixel.
For each of these 60 parameter combinations, we performed 24 trials, each with randomly chosen input phases, shot noise and readout noise.

For the linearized least-squares estimator, the linearization point was taken to be a flat wavefront, so $\bc_0 = 0$, resulting in an estimate $\bc_\mathrm{LLS}$.
The starting point for the nonlinear estimator was $\bc_\mathrm{LLS}$, resulting in a new estimator $\bc_\mathrm{NLS}$.
Both the linearized and nonlinear estimations forced the estimated phases to have zero-mean, as explained in Sec.~\ref{sec: Estimation Procedures}.\ref{sec: zero-mean}.
The enforcement of zero-mean  removed the only zero (or small) singular value in the inversion problem (see Fig.~\ref{fig: singular values}), hence, there was no benefit to having a regularization parameter $\alpha > 0$, and all of the estimators employed no regularization.
Simulations not shown here confirmed that regularization (with the regularization matrix set to the identity, so $\bR = \mathbb{1}$) provided no benefit to the accuracy of the estimators.
(The only exception to this was cases at low Strehl ratio in which the unregularized estimator gain was negative.
In these cases, a large regularization parameter drove the estimate towards 0, making the gain less negative.)
The weight matrix $\bW$\ in the cost functions, Eqs.~(\ref{eq: CostLLS}) and (\ref{eq: CostNLS}), was set to the identity matrix.
(Due to the very low intensity levels in some pixels of the detector, weighting of the measurements by the inverse of the variance noise caused poor performance of the algorithms.
Finding an optimal weighting procedure was not explored.)

\begin{figure}[t!]
\begin{tabular}{l|r}
\includegraphics[height=70mm]{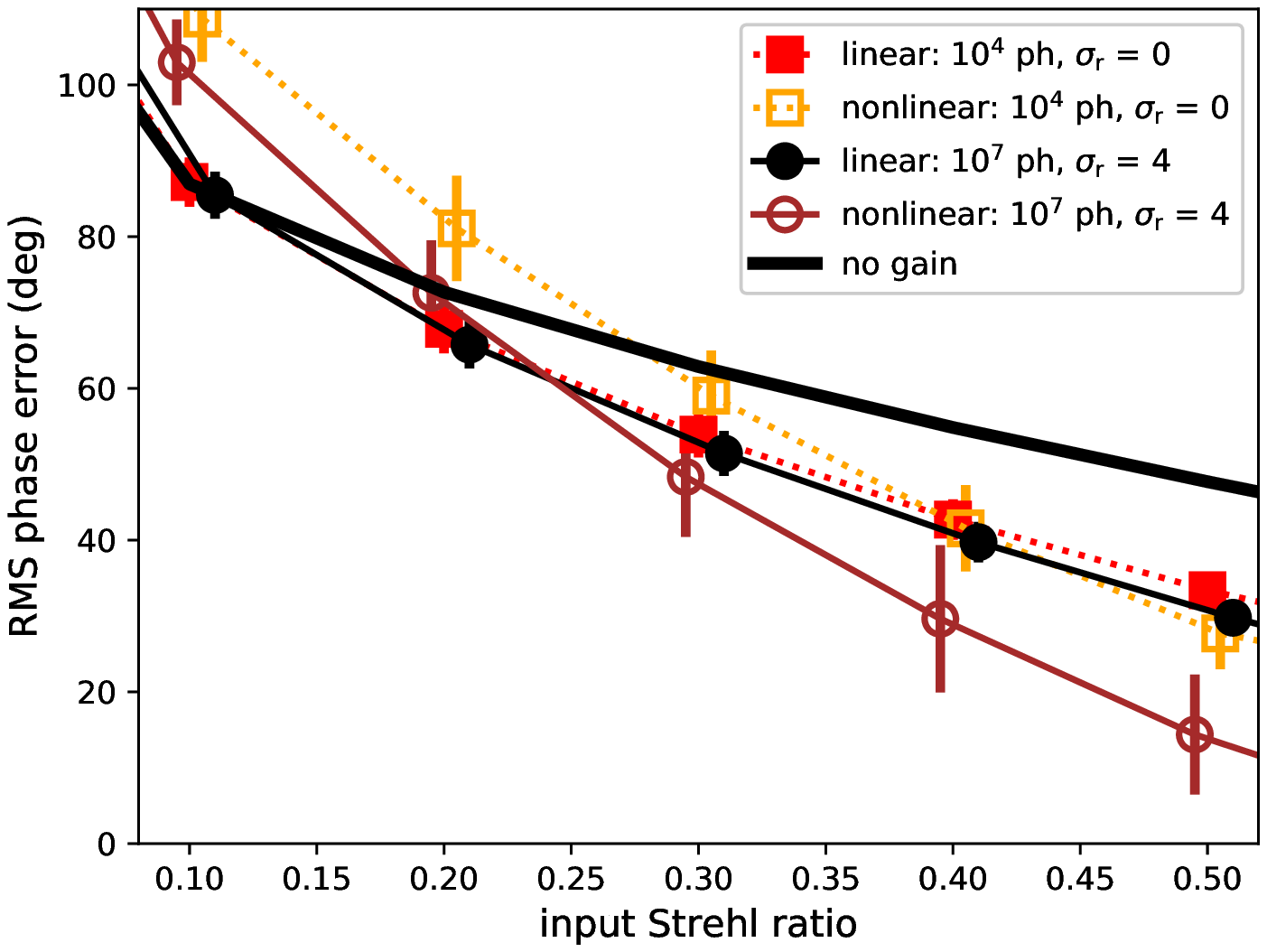} &
\includegraphics[height=70mm]{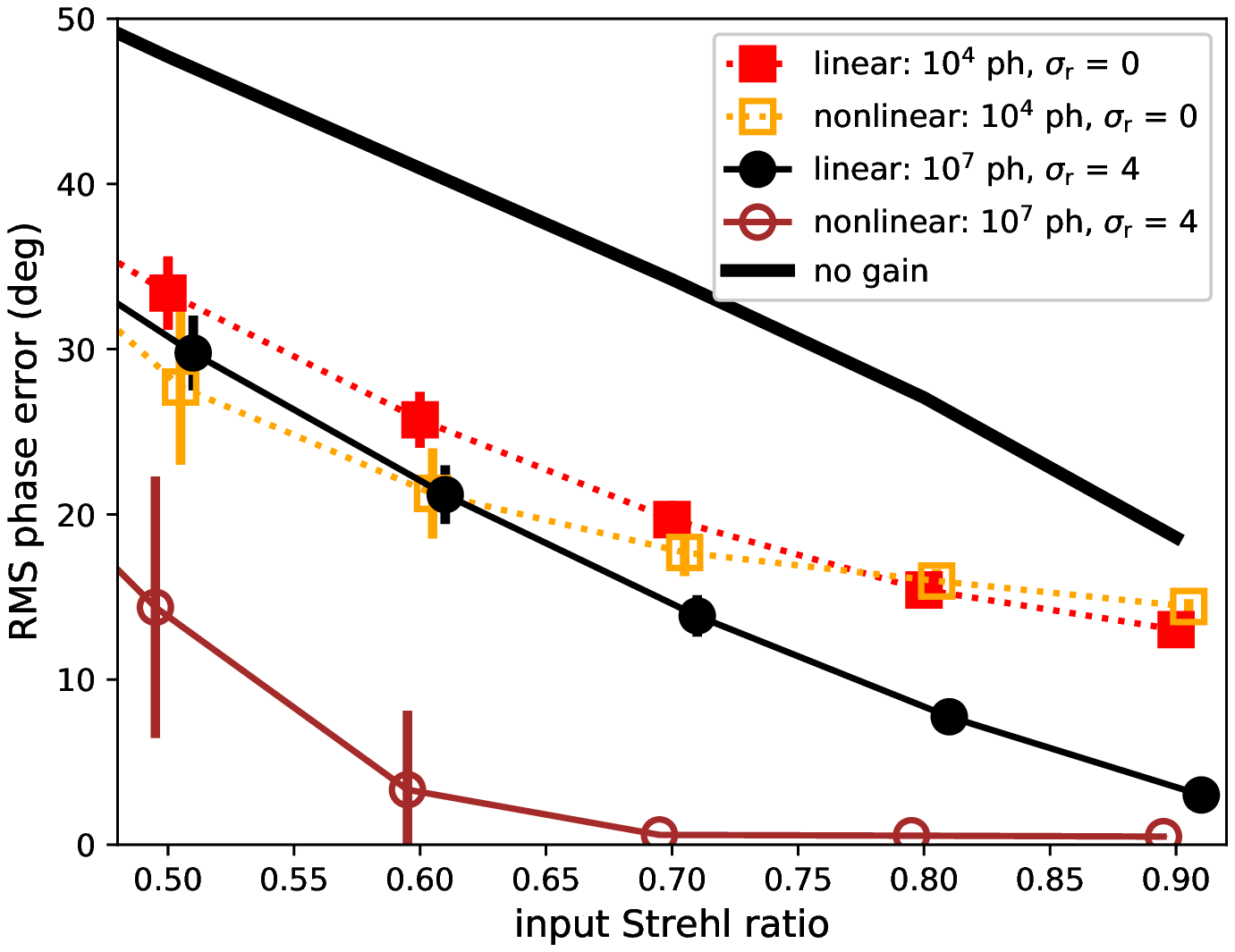} \\
\includegraphics[height=70mm]{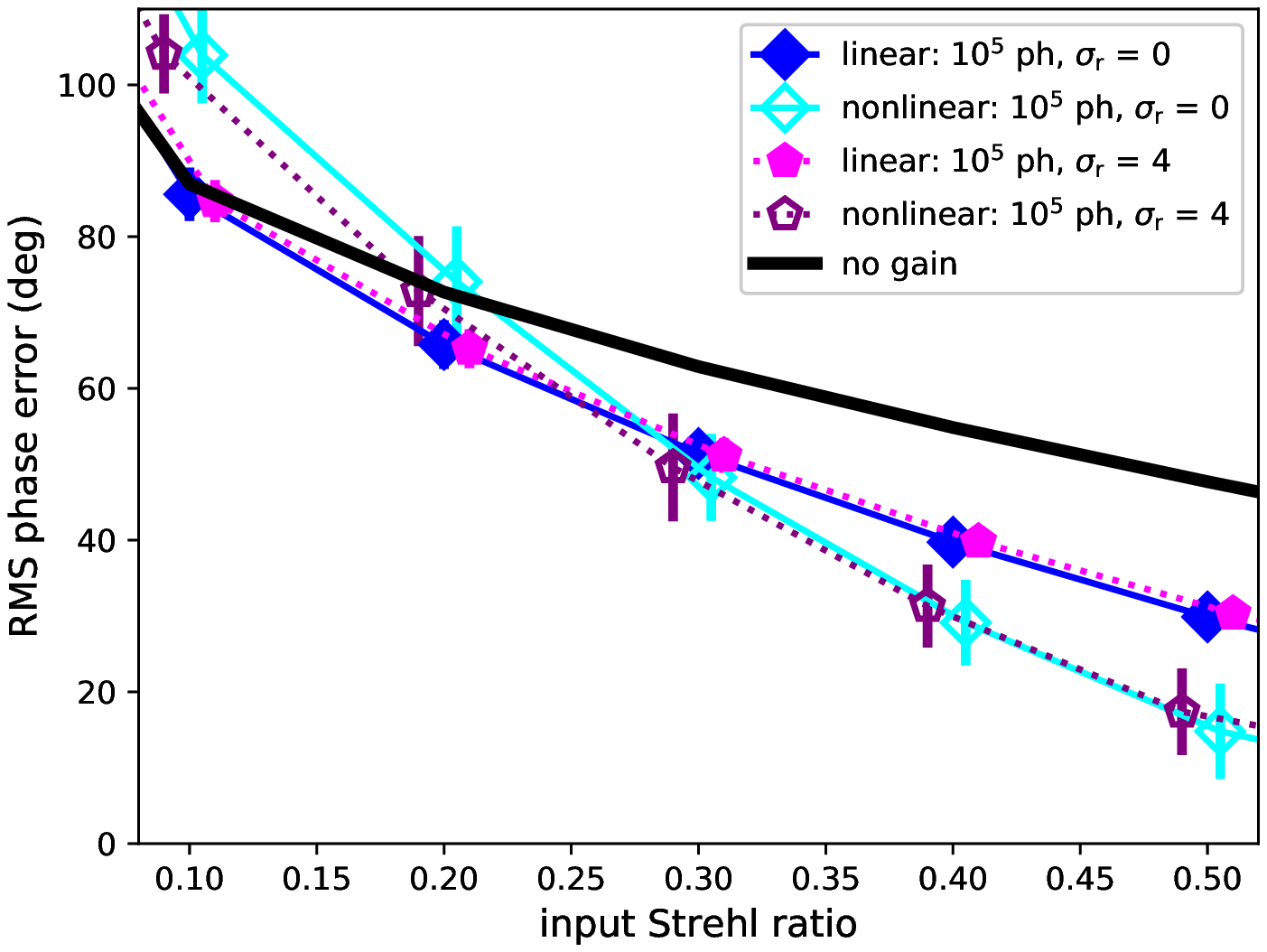} &
\includegraphics[height=70mm]{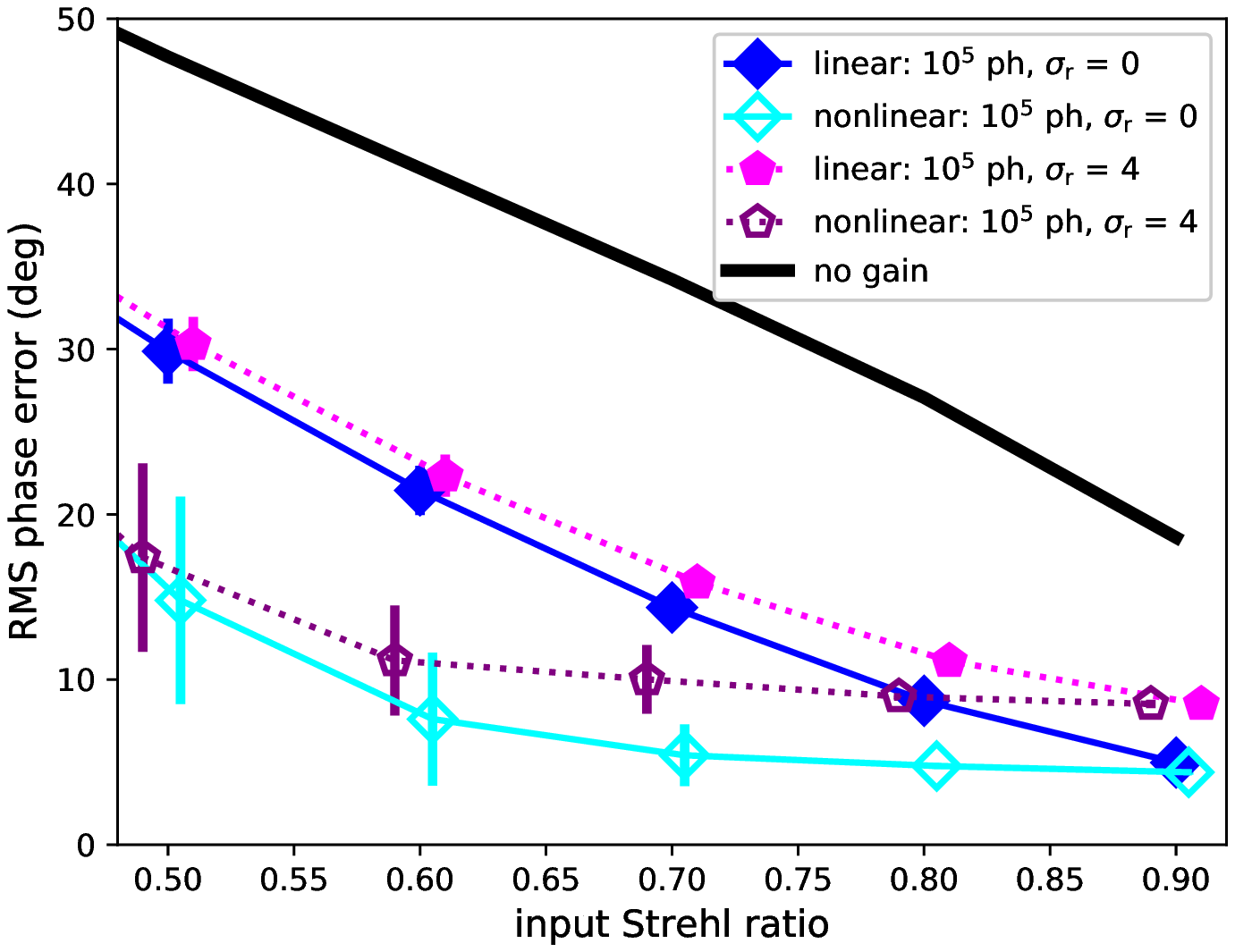}
\end{tabular}
\caption{\small Performances of linearized and nonlinear estimators.
The plots show the RMS error of the estimate of the phase on the $y$-axis, and the $x$-axis is the Strehl ratio of the input beam.
Except for the "no gain" curve, each curve corresponds to a different pairing of the number of photons incident on the detector and the detector readout noise level $\sigma_\mathrm{r}$.
Linearized and nonlinear estimator curves for the same photon count number and $\sigma_\mathrm{r}$ have the same line style and plotting symbol (except the plotting symbol for the linear estimator is filled, while the one for the nonlinear estimator is not) but have different colors. 
Each data point corresponds to the RMS error averaged over 24 trials, and the error bars represent the standard deviation of the RMS errors over the 24 trials.
Note that some error bars are too small to be seen. 
Small horizontal offsets are used to separate the data to make the plots easier to read.
\emph{upper plots:}
Curves for ($10^7$ photon counts, $\sigma_\mathrm{r} = 4$ counts/pixel), and for ($10^4$\ photons, $\sigma_\mathrm{r} = 0$).
The RMS error of the ($10^7$ photons, $\sigma_\mathrm{r} = 4$) curve is too small to be seen on the plot at Strehl ratios of 0.7, 0.8 and 0.9.
These values are $0.6^\circ, \, 0.55^\circ$, and $0.49^\circ$, respectively, and their standard deviations are all about $0.02^\circ$.
The curves for ($10^4$ photons, $\sigma_\mathrm{r} = 4$) showed no gain at any Strehl ratio and are not displayed here.
The curves for ($10^7$ photons, $\sigma_\mathrm{r} = 0$) are visually indistinguishable from the curves for ($10^7$ photons, $\sigma_\mathrm{r} = 4$), and are not displayed here.
\emph{lower plots:}
Curves for ($10^5$ photons, $\sigma_\mathrm{r} = 0$) and ($10^5$ photons, $\sigma_\mathrm{r} = 4$).
}
\label{fig: performances}
\end{figure}

Fig.~\ref{fig: performances} shows the results of the various runs.
The $y$-axis of plots displays the RMS error of the estimates calculated as:
\begin{equation}
\mathrm{RMS \; error} = \mathrm{std}( \mathrm{estimated \; phase} - \mathrm{true \; phase} ) \, ,
\label{eq: ErrorDef}
\end{equation}
where "std" signifies standard deviation, where standard deviation is carried out over the 797 phase values.
In each plot, the heavy black curve represents no gain, which is the error achieved by setting "estimated phase" to 0 in \eqref{eq: ErrorDef}, thus any estimator below the curve is providing some information.
Each data point shows the estimator error, according to \eqref{eq: ErrorDef}, averaged over the 24 trials for a given parameter set.
The data points have small horizontal offsets to alleviate visual over-crowding.
The error bars, when large enough to be visible, indicate the standard deviation of this error over the 24 trials.

No estimator we used provided gain below a Strehl ratio of 0.1, due to the large nonlinearity error.
This can be understood in terms of the amplification of nonlinear error upon inversion as explained in  \ref{sec: Estimation Procedures}\ref{sec: slope sensing} and Fig.~\ref{fig: nonlinear error}.
For example, with a Strehl ratio of 0.1, $10^7$ photon counts and $\sigma_\mathrm{r} = 0$, the linearized estimator only provided a gain of $\approx 1.5 \%$.
In this case, subsequently applying the BFGS algorithm resulted in a greatly decreased cost function.
[In this case  $\mathcal{C}_\mathrm{ l}(\bc_\mathrm{LLS}) / \mathcal{C}_\mathrm{nl}(\bc_\mathrm{NLS}) \approx 6$, but the gain achieved by the nonlinear estimate was $\approx - 18 \%$.]

Varying the number of photons showed that $10^4$ photons is the lowest intensity level in which the regressions provided any gain (with zero readout noise) and represents something close to the lower limit at which the PyWFS simulated here can hope to operate.
This seems reasonable, since ($10^4$ photons)/(797 pupil pixels)$~ \approx 12.5~$photons/pixel corresponds to a signal-to-noise ratio of about $12.5/\sqrt{12.5} = \sqrt{12.5} \approx 3.5$.
With a photon count of $10^4$ and $\sigma_\mathrm{r}=  4$ counts/pixel, neither the linearized nor nonlinear estimator showed gain, and the results for these runs are not displayed.

The top panels in Fig.~\ref{fig: performances} show the results for photon counts of $10^7$ and $10^4$, which are the lowest and the highest count rates we evaluated.
For $10^7$ photon counts, the curves for readout noise levels of $\sigma_\mathrm{r} = 4$ and $0$ counts/pixel are visually indistinguishable, so only the results for 4 counts/pixels are displayed.
For the case of the $10^7$ photons, $\sigma_\mathrm{r} = 4$, the RMS error of the nonlinear estimator is too small to be seen on the plot at Strehl ratios of 0.7, 0.8 and 0.9.
These values are $0.6^\circ, \, 0.55^\circ$, and $0.49^\circ$, respectively, and the standard deviations of the RMS errors over the 24 runs are all about $0.02^\circ$.
These plots show that for $10^7$ photons, the nonlinear estimator begins to have an advantage over the linear  estimator at a Strehl ratio of 0.3.
The advantage continues to grow (the ratio of linear estimator error to the nonlinear estimator error increases) as the Strehl ratio increases.
In contrast, the case of $10^4$ photons has such poor signal-to-noise, that there is little advantage to using the nonlinear estimator at any Strehl ratio.

The bottom panels in Fig.~\ref{fig: performances} show the results for $10^5$ photon counts and readout noise levels of 0 and 4 counts/pixel.
The effect of the readout noise levels is clearly visible in the lower-right panel, where the Strehl ratio is greater than 0.5.
In addition, the advantage of the nonlinear estimator over the linear one begins at a Strehl ratio of about 0.4.
Unlike the case with $10^7$ photons, when only $10^5$ photons are available, the linear estimator performs almost as well as the nonlinear one at a Strehl ratio of 0.9.
A similar effect can be seen in the upper panel for the $10^4$ photons case, were the linear estimator slightly outperforms the nonlinear estimator at a Strehl ratio of 0.9 (this was confirmed with several batches of 24 trials).
The cause of this is the fact that the nonlinear cost function has many local minima and, when the data are noisy, the minimization algorithm finds a minimum that is not more desirable than the linearized least-squares solution. 
Why this effect is more pronounced at a Strehl ratio of 0.9 than Strehl ratios 0.3 through 0.8 is a bit of a curiosity.
It is also interesting to note that performance of the estimators at Strehl ratios less than about 0.5 depends only weakly on the shot noise and detector noise.
This is because the error in the solution is dominated by the nonlinearity, which is only partially mitigated by the nonlinear estimation method.
At Strehl ratios of about 0.5 and above, the nonlinearity error is smaller, and the effects of noise are more pronounced.

\section{Conclusions/Discussion}

The desire to use the PyWFS without modulation stems from the high sensitivity of the device, but the price of this high sensitivity is nonlinearity.
The simulation experiments here indicate that when the Strehl ratio of the input beam is above some threshold and the SNR is sufficient, nonlinear estimation methods can provide substantial advantages over linear estimation methods, thereby improving prospects for using the PyWFS without modulation.
For the PyWFS simulated here, taking advantage of nonlinear estimation requires the input beam to have a Strehl ratio of about 0.3 or greater and photon flux about an order of magnitude greater than the operation threshold of the instrument.
We have also shown that traditional slope-based measurements, in which the four PyWFS pupil images are reduced to two slope images, are unsuited to using the PYWFS without modulation.
This is because the Jacobians associated with the slope representations of the measurements have steeper spectra of singular values than the Jacobians associated with the non-slope representations of the measurements.
The effect of a steeper spectrum of singular values is more amplification of nonlinearity error and noise upon inversion, as explained in Sec.~\ref{sec: Estimation Procedures}\ref{sec: slope sensing}

We have argued that employing nonlinear estimators based on Newton and quasi-Newton methods for optimization should be possible for AO systems running at kHz rates for two reasons.
Firstly, all detailed optical modeling is done beforehand, so no computationally expensive operations such as FFTs are required in real-time.
With no real-time constraints on the optical modeling, one can apply almost any computational modeling method. 
Secondly, the computations required to obtain the modeled intensity and its derivatives can be carried out efficiently and in a massively parallel manner, as explained in Sec.~\ref{sec: discretization}.
Using a computationally advantageous representation of the wavefront such as PhasePixel, as opposed to ModalPhase, reduces the real-time computational burden.
Table~\ref{table: CompCount} gives an approximate count of the required operations to calculate the intensity and its Hessian matrix at every detector pixel. 
For example, if specifying the model requires $N=10^3$\ phase values and there are $10^4$ detector pixels at which the intensity and its derivatives are to be determined, calculating the intensity requires $\sim \, 5 NL = 5 \times 10^7$ FLOPS, and calculating the intensity Jacobian requires an additional $NL = 10^7$ FLOPS.
Calculating the $N \times N$ Hessian matrix at every pixel is more expensive and requires an additional $\sim N^2L/2 = 5 \times 10^9$ FLOPS, but there may be possibilities to reduce this by approximating these matrices with sparse alternatives, as explained in Sec.~\ref{sec: discretization}\ref{sec: sparsity}.
In addition, every iteration of the optimization method will require more operations, as described in Sec.~\ref{sec: Estimation Procedures}\ref{sec: NLS}.
For perspective, the NVIDIA Tesla P100 GPU Accelerator delivers 10 TeraFLOPs ($10^{13}$) of single-precision arithmetic per second, so a highly streamlined system with multiple GPUs may be capable of the required real-time computations.

This article has emphasized the nonlinearity of the non-modulated PyWFS and the fundamental challenges it imposes on estimation of the wavefront.
The AO community has long recognized this challenge, and has used modulation of the input beam to make the PyWFS more linear.
In order to mitigate some of the nonlinearity while retaining sensitivity, the PyWFS could be used with a small amount of modulation, and nonlinear estimation could be applied in such a situation as well.
One obstacle to utilizing the estimation methods described here with a modulated PyWFS is increased computational expense of calculating the intensity and its derivatives.
Let us take circular modulation applied by a steering mirror as an example.
The tip/tilt applied by the steering mirror would need to be included in \eqref{eq: PhasePixelDetField} and the equations that follow, until the intensity and its needed derivatives are determined.
This would need to be done for multiple values of tip/tilt modulation in order to sample the path of the steering mirror.
Then the intensities (and derivatives) would be averaged.
Sampling the steering circle every 3.6 degrees would result in 100 times the computation of a calculation performed without modulation.
One way that may reduce this burden would be to apply to use a spatial light modulator with, say, 10 modulation settings, in lieu of beam steering.\cite{Wang_Py_SLM_OE11, Akondi_Py_SLM_13}

In both the modulated and non-modulated cases, determining the matrix $ \{ P\big( \be_k ; \, \bs_l \big) \}$, which is essentially a phase-accurate, calibrated computational model of the PyWFS, will be a challenge.
Such an effort would likely require a combination of calibration measurements, and simulated propagation.

The impact of the nonlinear estimation methods described here on real-time, closed-loop control requires more detailed investigation.
The next logical step would be end-to-end simulation in which the beam feeding the non-modulated PyWFS has been partially corrected by an upstream AO system.

The role of nonlinear estimation in post-analysis, where the time constraints are not as stringent, has not been mentioned.
If data are acquired with sufficiently high Strehl ratio and signal-to-noise, this nonlinear estimation here should produce accurate estimates of the pupil phases and possibly the amplitudes as well.
This capability would enable millisecond approaches to high-contrast imaging in which the wavefronts are needed as inputs to regression schemes.
One such regression method produces simultaneous and self-consistent estimates of both the exoplanet image and the non-common path aberrations in the optics.\cite{Frazin13}

\section*{Acknowledgments}
The author thanks the referees, Olivier Guyon, and John Kohl for comments on the manuscript.
This work has been supported by NSF Award \#1600138 to the University of Michigan. 

\clearpage

\clearpage


\end{document}